\documentclass{article}

\usepackage[english]{babel}

\usepackage[letterpaper,top=2cm,bottom=2cm,left=3cm,right=3cm,marginparwidth=1.75cm]{geometry}

 \usepackage{multirow}
\usepackage{amsmath}
\usepackage{amsmath, amsthm, amssymb, amsfonts}
\usepackage{mathtools}
\usepackage{graphicx}
\usepackage{xcolor}
\usepackage[colorlinks=true, allcolors=blue]{hyperref}
\usepackage{tensor}
\usepackage{wrapfig}
\usepackage{caption}
\usepackage{subcaption}
\usepackage{cite}
\usepackage{url}

\def\ee{\end{equation}}
\def\eea{\end{eqnarray}}
\def\be{\begin{equation}}
\def\bea{\begin{eqnarray}}
\def\bal{\begin{aligned}}
\def\eal{\end{aligned}}
\def\p{\partial}
\def\AB{(A \leftrightarrow B)  }
\def\mphi{m_{\varphi}}
\title{Dynamics of compact binary systems in massive scalar Gauss-Bonnet gravity}

\begin{document}
\maketitle
\begin{center}\textbf{
Iris van Gemeren\textsuperscript{$\star$},
Tanja Hinderer\textsuperscript{$\dagger$} and
Stefan Vandoren\textsuperscript{$\ddagger$}
}\end{center}

\begin{center}
Institute for Theoretical Physics,
Utrecht University, Princetonplein 5, 3584 CC Utrecht, The Netherlands
\\[\baselineskip]
$\star$ \href{mailto:email1}{\small i.r.vangemeren@uu.nl}\,,\quad
$\dagger$ \href{mailto:email2}{\small t.p.hinderer@uu.nl}\,,\quad
$\ddagger$ \href{mailto:email2}{\small s.j.g.vandoren@uu.nl}
\end{center}

\section*{Abstract}
Inspiraling binary systems of compact objects probe gravity in strong-field regimes, thereby exploring potential higher curvature corrections to General Relativity. Parity-invariant quadratic corrections can be described by scalar-Gauss-Bonnet (sGB) theory, which involves a scalar field dynamically coupled to curvature scalars and can give rise to scalar condensates around black holes. Considering a mass for the scalar field is natural and leads to new phenomenology related to this additional scale. We compute the dynamics of a binary system of nonspinning black holes in massive sGB using the post-Newtonian (PN) approximation. We obtain solutions  
for the equations of motion, center-of-mass transformation, and binding energy for circular and eccentric orbits up to 1PN order, where for the first time the higher curvature coupled to scalar mass corrections are included.
While most of our calculations are valid for generic scalar masses, the final explicit expressions assume that the mass is small compared to the total mass of the binary and expand to quadratic order in this ratio. We show that the scalar mass corrections to the gauge-invariant binding energy come with same and opposite sign order terms, contributing an overall opposite sign contribution in the perturbative limit, decreasing the binding energy slightly. The effects are largest for binary systems with high mass ratio and large eccentricity. Our methods and results will also be useful as a basis for computing the gravitational waves sourced by such systems. 

\tableofcontents
\clearpage

\section{Introduction}

Gravitational waves (GWs) generated by binary systems probe the strong gravity regime around compact objects, allowing for unprecedented tests of General Relativity (GR)~\cite{LIGOScientific:2016lio,LIGOScientific:2019fpa,LIGOScientific:2020tif,LIGOScientific:2021sio,LIGOScientific:2026qni,LIGOScientific:2026oim}. In particular, GWs probe potential beyond-GR effects that become relevant only when the gravitational curvature is strong enough. Such modifications can easily satisfy constraints from empirical tests of gravity~\cite{Will:2014kxa,Turyshev:2008dr,Berti:2015itd} such as tabletop experiments~\cite{Adelberger:2003zx}, solar system tests~\cite{Will:2014kxa} and constraints from binary pulsars~\cite{Wex:2014nva}, while GW observations allow for placing tighter constraints. For GWs, extracting the information from the data requires accurate theoretical models that include beyond-GR modifications. While consistency tests of GR can be carried out with parameterized deviations from GR waveforms, linking the GW data to constraints on fundamental theories requires waveforms in example classes of theories beyond GR. Of particular interest for probing higher-curvature corrections is the inspiral epoch of a binary coalescence, as the progenitors have smaller masses and hence higher curvatures than the remnant black hole and differences in the phase evolution of the GWs with respect to GR can accumulate over many cycles, potentially leading to net observable changes. In this regime, analytical approximations that exploit the hierarchy of scales in the system at large separation can be used to compute the binary dynamics and waveforms~\cite{Blanchet:2013haa}. In particular, the early inspiral is well-approximated by Post-Newtonian (PN) theory, which assumes both weak fields and low velocities. 

In this paper we compute the dynamics of binary systems in scalar-Gauss-Bonnet (sGB) theories~\cite{Kanti:1995vq} with a massive scalar field in the PN approximation. In general, sGB theories involve an additional scalar field non-minimally coupled to the quadratic-in-curvature Gauss-Bonnet invariant. The coupling to the Gauss-Bonnet topological invariant term ensures the theory is ghost-free~\cite{Nojiri:2018ouv} and still second order in the field equations. The sGB corrections to the Hilbert-Einstein action also have motivations from the low energy limit of quantum gravity paradigms~\cite{Kanti:1995vq, Zwiebach:1985uq, Gross:1986mw, Boulware:1985wk}. For massless scalar fields, sGB gravity is well studied, for instance, depending on the parameters of the theory, black holes can evade the no-hair theorem. For linear or dilatonic coupling functions the black hole solutions in sGB always differ from GR, being dressed with a scalar profile 
extending from within the horizon~\cite{R:2022tqa, Kanti:1995vq, Pani:2009wy, Sotiriou:2014pfa, Benkel:2016rlz, Antoniou:2017acq,Antoniou:2017hxj, Papageorgiou:2022umj, Prabhu:2018aun, Saravani:2019xwx, Ripley:2019irj, Sotiriou:2013qea, Sullivan:2019vyi, Ayzenberg:2014aka, Maselli:2015tta,Kleihaus:2011tg, Kleihaus:2014lba,Kleihaus:2015aje}. For quadratic or Gaussian coupling functions the scalarization or descalarization and thus the deviation from GR black holes or neutron stars can arise spontaneously~\cite{Silva:2017uqg, Dima:2020yac, Herdeiro:2020wei, Berti:2020kgk, Collodel:2019kkx, Doneva:2020nbb, Doneva:2017bvd, Cunha:2019dwb}. See the review articles~\cite{Doneva:2022ewd,Herdeiro:2015waa} for an extensive overview. For binary systems in the regime of weak sGB couplings, the theory was shown to be well-posed~\cite{Kovacs:2020pns,Kovacs:2020ywu,AresteSalo:2022hua}. 

Under the assumption that the sGB scalar field is massless, the effects on the binary dynamics and waveforms on the inspiral are modeled up to first order in the PN expansion~\cite{Julie:2019sab,Julie:2019sabErr,Shiralilou:2020gah,Shiralilou:2021mfl} including leading order dipolar tidal effects~\cite{vanGemeren:2023rhh}.
Numerical relativity computations of GWs in sGB in the weak-coupling approximations have also been performed~\cite{East:2020hgw,East:2021bqk,Corman:2022xqg,Witek:2020uzz, Elley:2022ept,Silva:2020omi, Doneva:2022byd, Corman:2025wun, Capuano:2026lhs, AresteSalo:2025sxc, East:2022rqi,Corman:2024vlk, AresteSalo:2026zzc, Corman:2024cdr, AresteSalo:2022hua}. Together with studies of the quasinormal mode frequencies and ringdown signature of black holes in sGB gravity~\cite{Bryant:2021xdh, Hu:2025bkg, Blazquez-Salcedo:2017txk, Chung:2024ira, Chung:2024vaf, Pierini:2022eim, Blazquez-Salcedo:2024oek}, an inspiral-merger-ringdown effective one body model was recently built for sGB gravity~\cite{Julie:2024fwy}. 
Based on all of the above progress, the tightest constraints on the massless theory for the coupling constant $\alpha$ of the Gauss-Bonnet invariant to the scalar field are coming from GW observations~\cite{Witek:2018dmd, Nair:2019iur, Perkins:2021mhb, Wang:2021jfc,Wang:2023wgv,Gao:2024rel,Lyu:2022gdr,Julie:2024fwy},
currently around $\sqrt\alpha\lesssim 0.31\textrm{km}$~\cite{Julie:2024fwy}.

The massive scalar field extension of sGB theory is natural from a theoretical point of view and corresponds to the leading order self-interaction of the scalar. Massive scalars can be found more broadly in physics, including the only measured scalar field to this day: the Higgs boson, and proposed dark matter candidates as axion-like particles and other ultralight boson models~\cite{Peccei:1977hh,Hui:2016ltb,Arvanitaki:2010sy,Ferreira:2020fam}. A main feature of scalarized black holes in massive sGB is that the additional lenghtscale introduced by the mass term exponentially suppresses the scalar field for scales larger than its Compton wavelength~\cite{vanGemeren:2024bzf}, whereas in the massless limit, it falls off inversely with distance. Thereby the massive theory automatically evades constraints on larger scales. 
Black holes with massive scalar configuration have been studied numerically~\cite{Doneva:2019vuh} and perturbatively for static~\cite{vanGemeren:2024bzf}  and rotating bodies~\cite{Chung:2024ira}, and the onset for spontaneous scalarization analysed~\cite{Macedo:2019sem}. Furthermore black holes in the large mass regime~\cite{Hod:2019vut}, with cosmological constant~\cite{Bakopoulos:2020dfg} and their axial quasinormal modes have been considered. Spontaneously scalarized neutron stars have been studied as well~\cite{Xu:2021kfh}. The current constraints on the coupling constant $\alpha$ for large scalar masses come from theoretical arguments~\cite{vanGemeren:2024bzf}, while for small masses they are set by GW observations~\cite{Yamada:2019zrb, Xie:2024xex} currently bounding it to be $\sqrt{\alpha}\lesssim 1 \textrm{km}$ for scalar masses $m_{\varphi}\lesssim 10^{-13}\textrm{eV}$.

More broadly, massive scalar fields in the context of compact objects are an active area of study, for instance, massive dilaton fields coupled to charged black holes~\cite{Horne:1992bi, Boyadjiev:2002en}, in the context of black hole superradiance as reviewed in~\cite{Brito:2015oca},  and compact objects in massive scalar-tensor theories~\cite{Staykov:2018hhc, Ramazanoglu:2016kul, Yazadjiev:2016pcb, M:2026whf, Hu:2021tyw, Degollado:2024oyo} or dynamical-Chern-Simons gravity~\cite{Macedo:2018txb}. On the observational side the effects of a massive scalar have been studied in the light of searches for ultralight dark matter with LIGO and LISA detectors~\cite{Brito:2017zvb}, and future probes of massive scalar fields via multiband detections~\cite{Chen:2024ery} or  extreme mass ratio systems~\cite{Barsanti:2022vvl,Zi:2026cwm, Li:2025ffh}. Furthermore the gravitational radiation from the inspiral has been modeled for the case of massive Brans-Dicke theory~\cite{Alsing:2011er,Liu:2020moh}, scalar tensor theories~\cite{Diedrichs:2023foj} and massive Horndeski models~\cite{Ozer:2025muv}. Lastly the quasinormal modes and the late time tail in the presence of massive scalar fields have been considered~\cite{Shao:2026hnd, Dubinsky:2024jqi, Burko:2004jn, Konoplya:2013rxa}.

In this work we explore the effects of the scalar field mass and higher curvature coupling corrections on the binary dynamics to first order in the Post-Newtonian expansion. This develops the methodology and is the first step for computing the GW signals and inspiral evolution. We obtain, for the first time, the scalar mass-higher curvature coupling corrections to the equations of motion, center-of-mass transformations and binding energy. Additionally we investigate how the direct integration approach to solving the PN equations~\cite{Pati:2000vt} must be adapted for massive sGB gravity, showing that the scalar mass introduces nontrivial changes.
We end with a systematic parameter space study of the corrections to the binding energy as a function of frequency or constants characterizing the orbital geometry for circular and eccentric orbits. 

The organization of this paper is as follows, in Sec.~\ref{sec:msGB} we introduce the massive sGB action, field equations and we give an overview of the different scales in the system. In Sec.~\ref{sec:fieldsol} we compute the solutions for the gravitational and scalar field using the PN and weak field expansions. With the 1PN field solutions at hand, we obtain the equations of motion, center-of-mass transformations and binding energy up to the same PN order in Sec.~\ref{sec:BinaryDyn}. At the end of this section we study the dependence of the binding energy on the orbital frequency and orbital parameters for circular and eccentric orbits for different mass ratio choices and we conclude in Sec.~\ref{sec:conclusion}.\\ 
\emph{ Notation}. We use Greek indices to denote components on four-dimensional spacetime, Latin indices for three-dimensional spatial components, and boldface to indicate spatial vectors. Superscripts between round brackets $x^{(01)}$ denote PN order as a first index and order in the scalar field mass as the second index. A capital letter superscript denotes a product of that dimensionality $x^L \equiv x^{k_1} x^{k_2} \cdots x^{k_l}$.

\section{Massive scalar-Gauss-Bonnet gravity}\label{sec:msGB}

\subsection{Action and equations of motion of massive sGB theory}

Scalar-Gauss-Bonnet gravity is defined by the action

\be\label{eq:SmsGB}
S=  \frac{c^4}{16 \pi G} \int d^4 x \sqrt{-g}\left(R-2 g^{\mu \nu} \nabla_\mu \varphi \nabla_{\nu} \varphi-2 m_{\varphi}^2 \varphi^2+\alpha f(\varphi) R_{G B}^2\right) \\
+S_{\rm mat}(\Psi_{\rm mat}, \mathcal{A}^2(\varphi)g_{\mu\nu})
\ee
Here $\varphi$ is the additional scalar field with mass parameter $m_{\varphi}$ having dimension of inverse length. $m_{\varphi}$ is related to the scalar field mass in kilograms $m_s$ via $m_{\varphi}=\frac{m_s c}{\hbar}$. When, in the rest of this paper, we refer to the scalar field mass, we refer to $\mphi$ not $m_s$. The scalar field is coupled via coupling constant $\alpha$ and coupling function $f(\varphi)$ to the Gauss-Bonnet invariant $\mathcal{R}_{G B}^2=R^2-4 R^{\mu \nu} R_{\mu \nu}+R^{\mu \nu \rho \sigma} R_{\mu \nu \rho \sigma}$. We keep the coupling function general until our final analysis. $S_{mat}$ is the matter action depending on the matter fields $\Psi_{mat}$ which are nonminimally coupled to the metric via the function $\mathcal{A}(\varphi)$.\\
The action is expressed in the Einstein frame as the scalar field is not coupled to the Ricci scalar and the GR limit is preserved. The coupling of the scalar field to the metric happens via the matter action. Via a conformal transformation the scalar field can become coupled to the Ricci scalar and uncoupled to the matter sector. This frame is called the Jordan frame which thus leaves the matter sector unaffected. Deriving the Gauss-Bonnet corrections from the low energy limit of quantum gravity theories such as string theories, the action is obtained in the Jordan frame with $A(\varphi)=e^{\varphi}$ for dilatonic couplings. A different approach is to see~\eqref{eq:SmsGB} as an effective action which includes higher order curvature corrections to GR. From this perspective the Einstein frame formulation is the fundamental frame. One expects the physical observables not to depend on the choice of frame. Additionally, in the weak coupling limit, the scalar field is small and the conformal factor can be approximated to be 1. Throughout this article we work in the Einstein frame formulation of the action as presented in~\eqref{eq:SmsGB}.

Varying the action~\eqref{eq:SmsGB} with respect to the metric and scalar fields leads to the following field equations
\begin{subequations}\label{eq:FEfull}
\begin{eqnarray}
G^{\mu \nu}&=&\frac{8 \pi G}{c^4} T_{\rm mat}^{\mu \nu}+2 \nabla^\mu \varphi \nabla^{\nu}\varphi-g^{\mu \nu}(\nabla \varphi)^2+g^{\mu \nu} m_{\varphi}^2 \varphi^2  -4 \alpha\left({ }^*R^{*^{c \mu \nu d}} \nabla_c \nabla_d f(\varphi)\right), \label{eq:EFE1}\\
\left(\square_g-\mphi^2\right) \varphi&=&-\frac{1}{4} \alpha f^{\prime}(\varphi) R_{G B}^2+\frac{4 \pi G}{c^4} \frac{1}{\sqrt{-g}} \frac{\delta S_{\rm mat}}{\delta \varphi},\label{eq:scalareom1}
\end{eqnarray}
\end{subequations}
with $\square_{g}=g_{\mu\nu}\nabla^{\mu}\nabla^{\nu}$ the curved space d'Alembertian. 

\subsection{Reformulation of the field equations}
For later use, it is convenient to introduce the change of variable
\begin{equation}
h^{\mu\nu}=\eta^{\mu\nu}-\sqrt{-g} \, g^{\mu \nu}
\end{equation}
and the harmonic gauge condition $\p_{\mu}(\sqrt{-g} \, g^{\mu \nu})=0$, which is equivalent to 
\begin{equation}\partial_{\beta}h^{\alpha\beta}=0.\label{eq:harmonicgauge}
\end{equation}
The field equations~\eqref{eq:EFE1} can then be rewritten as wave type equations
\begin{subequations}\label{eq:FEh}
\be
\square_{\eta} h^{\alpha \beta} = -\frac{16 \pi G}{c^4} \mu^{\alpha\beta}, \label{eq:Boxh}
\ee
with
\be\label{eq:sourcetensor}
\begin{aligned}
\mu^{\alpha \beta}=&(-g) T_{\rm mat}^{\alpha \beta}  +(-g) t_{L L}^{\alpha \beta}+\frac{c^4}{16{\pi} G}\left(\partial_\mu h^{\alpha \nu} \partial_\nu h^{\beta \mu}-h^{\mu \nu} \partial_{\mu \nu} h^{\alpha \beta}\right) \\
& +\frac{c^4}{16 \pi G}\bigg[4 \partial_{\rho} \varphi \partial_\sigma \varphi\left(\bar{g}^{\alpha {\rho}} \bar{g}{ }^{\beta \sigma}-\frac{1}{2} \bar{g}^{\alpha \beta} \bar{g}^{{\rho} \sigma}\right) -8 \alpha(-g)\left({}^*\hat{R}^{*^{ {\rho}\alpha \beta \sigma}} \nabla_\rho \nabla_\sigma f(\varphi)\right) \\
&-\bar{g}^{\alpha \beta} m_{\varphi}^2 \varphi^2\bigg].
\end{aligned}
\ee
\end{subequations}
Here $t_{LL}^{\alpha\beta}$ it the Landau-Liftshitsz pseudotensor~\cite{Landau:1975pou}, $\bar{g}^{\mu \nu}=\sqrt{-g} \, g^{\mu \nu}$ and $\square_{\eta}=\eta^{\mu\nu}\partial_{\mu}\partial_{\nu}$ the flat space d'Alembertian. 

We can express the scalar field equation of motion~\eqref{eq:scalareom1} as
\be\label{eq:FEphi}
(\square_g -\mphi^2)\varphi = \frac{4 \pi G}{c^4} \mu_s,
\ee
with
\be\label{eq:sourcescalar}
\mu_s=\frac{1}{\sqrt{-g}} \frac{\delta S_{\rm mat}}{\delta \varphi}-\frac{c^4}{16 \pi G} \alpha f(\varphi) \hat{R}_{G B}^2.
\ee
 The hatted quadratic curvature terms are the gauge fixed GB and double dual Riemann tensors, see Appendix A of~\cite{Shiralilou:2021mfl} for the explicit expressions. Together with the harmonic gauge condition~\eqref{eq:harmonicgauge}, 
the system \eqref{eq:FEh} and \eqref{eq:FEphi} are an exact representation of the modified Einstein and scalar field equations~\eqref{eq:FEfull}. Equivalently the gauge condition can be stated as the conservation equation $\partial_{\beta}\mu^{\alpha\beta}=0$.

We can further rewrite the scalar field equation to explicitly involve the d'Alembertian for flat spacetime using $$\square_g\varphi = \frac{1}{\sqrt{-g}}(\partial_{\nu}(\sqrt{-g}g^{\mu\nu})\partial_{\mu} + \sqrt{-g}g^{\mu\nu}\partial_{\mu}\partial_{\nu})\varphi,$$ where the first term vanishes due to the harmonic gauge condition~\eqref{eq:harmonicgauge}. Then we can rewrite the LHS of \eqref{eq:FEphi} as
$$(g^{\mu\nu}\partial_{\mu}\partial_{\nu}-m_{\varphi}^2)\varphi = (\frac{1}{\sqrt{-g}}(\eta^{\mu\nu}-h^{\mu\nu})\partial_{\mu}\partial_{\nu} - m_{\varphi}^2)\varphi.$$ The scalar field equation \eqref{eq:FEphi} then becomes
\be
\label{eq:finaleomphi}
\begin{aligned}
(\square_{\eta}-\sqrt{-g}m_{\varphi}^2)\varphi &= \frac{4 \pi G}{c^4} \sqrt{-g} \mu_s +h^{\mu\nu}\partial_\mu\partial_\nu\varphi.
\end{aligned}
\ee
So far, the derivations kept the matter source generic. We next specialize to a binary system at large separation to compute it explicitly. This requires first understanding the hierarchy of scales involved, which motivates corresponding approximations as we discuss below.  

 \subsection{Relevant scales in a binary system during inspiral}
\begin{figure}[h!]
\centering
\begin{subfigure}{1\textwidth}
   \includegraphics[width=\linewidth]{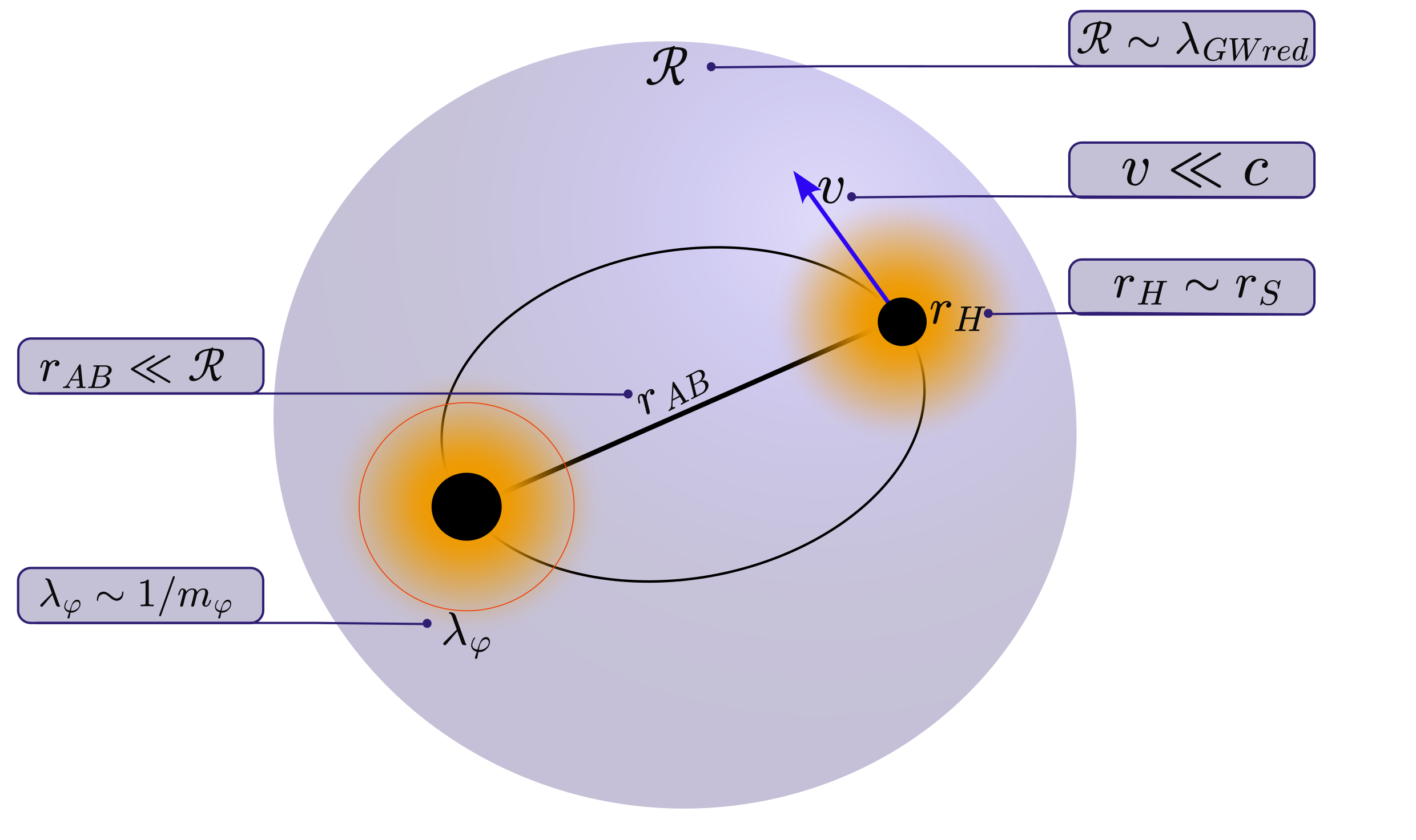}
       \centering
\end{subfigure}
\caption[]{Relevant scales and approximations in a binary system during inspiral.}
\label{fig:scales}
\end{figure}

 This work focuses on deriving the binary dynamics of a two-body system consisting of non-spinning black holes, labeled A and B, each dressed with a non-trivial scalar field cloud. We assume the binary is in its inspiral phase in which the separation $r_{AB}$ is large. Due to the large separation there is a clear distinction in length and time scales. We can assume the adiabatic approximation; the timescales for the source dynamics are much faster than the changes due to GW losses, hence we assume the system does not change during the orbital period. The system can be characterized by different length scales. We summarize the relevant scales of the system in Fig.~\ref{fig:scales}. The size of black holes in msGB theory is of the order of the Schwarzschild radius~\cite{vanGemeren:2024bzf}
\be
r_S = \frac{2 G m}{c^2},
\ee
with $m$ the mass of the black hole. The size of the scalar field around and inside the black hole is characterized by the Compton wavelength $\lambda_{\varphi}$ of the scalar field which is inversely proportional to the scalar field mass
\be
\lambda_{\varphi} \sim \frac{1}{\mphi}. 
\ee
Thus for small masses the scalar field stretches far outside the black holes while for very large masses the scalar field decouples from the body. During the inspiral we assume the separation to be much larger than the size of the black holes
\be
\label{eq:scalingrSvsrAB}
r_S \ll r_{AB},
\ee
such that effectively the black holes can be approximated as point particles with a scalar field dependent mass. The latter holds if also the scale of the scalar cloud is much smaller than the separation. A priori we would like to not set any constraints on the scalar field mass. However even in the massless limit it was shown that more than $90\%$ of the energy density of the field is concentrated in a region of a couple times the Schwarzschild radius~\cite{vanGemeren:2023rhh}. Hence the contribution from the scalar field outside of this range is assumed to be negligible. 
Independently the coupling constant $\alpha$ sets a lengthscale that needs to be determined by observational constraints. Currently for scalar field masses below $\mphi\lesssim 10^{-13} \textrm{eV}$ this is around $\sqrt{\alpha}\lesssim 1 \textrm{km}$. For that scalar mass range the coupling constant sets the smallest length scale in the system. For larger scalar field masses the tightest constraint comes from a theory argument~\cite{vanGemeren:2024bzf}. It's convenient to define dimensionless parameters related to the coupling constant and scalar field mass, rescaling them related to the total mass of the two bodies $M=m_A+m_B$
\be
\tilde{\alpha} = \frac{\alpha c^4}{4 G^2 M^2},
\ee
\be
\tilde{m}_{\varphi} = \frac{2 G M \mphi }{c^2}.
\ee
Note this definition is different from the dimensionless coupling and mass parameters defined in~\cite{vanGemeren:2024bzf} which are rescaled with the mass of an isolated black hole. In our calculations we don't assume the coupling constant to be small, however a small coupling is implicit when approaching \eqref{eq:SmsGB} as an effective description. For the scalar field mass, we keep the scale undefined up to the end of section~\ref{sec:1PNNZfields}. Afterwards we assume $\tilde{m}_{\varphi}\ll1$ and expand up to second order in the small scalar field mass. \\
The main interest as follow up work on this study is to compute the radiative moments of the fields and use energy balance to obtain the phase evolution of the gravitational radiation from the binary system. Alluding forward to this work we expect similar qualitative features as the leading and next to leading order gravitational power losses computed for massive scalar tensor theories in~\cite{Alsing:2011er, Diedrichs:2023foj}. It is found that the dipolar and quadrupolar scalar radiation gets activated when the orbital frequency is larger than the scalar field mass and half the scalar field mass respectively. The length scale of the scalar field mass thus plays a key role in the gravitational signature. Of observational interest are therefore the scalar field masses for which this activation happens during inspiral part that is picked up by the detectors.\\
Another large scale is the reduced wavelength $\lambda_{GWred}$ which distinguishes two scales at which different physics dominates, the near zone for scales smaller than the wavelength and the far zone for scales larger. The near zone boundary $\mathcal{R}$ is set by the wavelength scale
\be
\mathcal{R}\sim \lambda_{GWred}.
\ee
Within the near zone the separation between the bodies is large and the orbital velocity is not too large compared to the speed of light
\be
\label{eq:PNapprox}
\frac{G M}{r c^2}\sim\frac{v^2}{c^2}\ll1,
\ee
defining the so called Post-Newtonian expansion. In the far zone, retardation effects, considered small in the PN expansion, become important and only the weak field approximation holds. \\
The calculation in this paper is scheduled as follows. The solution for the gravitational and scalar fields are obtained at the scale of the binary system up to 1st order in the PN expansion. Additionally the contribution to the fields from the far zone are obtained, giving the full solution up to effective 1PN order. The total field solutions should not depend on the near zone boundary scale $\mathcal{R}$ which we show explicitly up to this order in the expansion. Subsequently we stay at the scale of the binary system to obtain the equations of motion and binding energy of the two body system, approximating the bodies as two point particles with the respective scalar dependent mass. We assume the scalar field mass to be small and expand the expressions to second order. From a short final analysis of the binding energy we analyze the dependence on the scalar field mass scale via the dimensionless mass parameter $\tilde{m}_{\varphi}$.

\subsection{Effective worldline description of scalarized black holes at large separation} 
In the regime of orbital separations that satisfy~\eqref{eq:scalingrSvsrAB}, the size of a scalarized black hole is small compared to the orbital scale and appears effectively as a point particle in a first approximation. This motivates approximating the full matter action~\eqref{eq:Spp} for a compact binary system at large separation by that of two point particles (for body A and B)  
\be\label{eq:Spp}
S_{\rm mat}\approx S_{p p}=-c \int M_A(\varphi) d s_A+\AB
\ee
In the point particle action, following the approach by Eardley~\cite{Eardley1975}, we capture the effect of the scalar field by introducing a scalar dependent mass $M_A(\varphi)$ of each body, and $ds_A=\sqrt{-g_{\mu\nu}dx_A^{\mu}dx_A^{\nu}}$ the differential along the worldline. This provides a coarse-grained description of the two scalarized black holes viewed on the large scales of the orbit.

Using the approximate matter action~\eqref{eq:Spp}, we compute explicit expressions for the matter-dependent source terms in the tensor~\eqref{eq:sourcetensor} and scalar~\eqref{eq:sourcescalar} field equations. This leads to
\be\label{eq:Tmunu}
\bal
T_{\rm mat}^{\mu \nu}&\approx\frac{1}{\sqrt{-g}} M_A(\varphi) \frac{u_A^{\mu} u_B^{\nu}}{u_A^0} \delta^3(x-x_A(t))+\AB, \\
\frac{\delta S_{\rm mat}}{\delta \varphi}&\approx\frac{1}{\sqrt{-g}} \frac{d M_A(\varphi)}{d \varphi} \sqrt{u_A^\mu u_{B_{\mu}}} \delta^3\left(x-x_A(t)\right) + \AB,
\eal
\ee
where the approximation is that the black holes are replaced by the effective worldline skeletons. 
Here ${ }^* R_{c \mu \nu d}^*=\frac{1}{4} \epsilon^{a b e f} R_{e f g h} \epsilon^{g h c d}$ is the double dual of the Riemann tensor and $u_A^{\mu}$ the four velocity. 

 \subsection{Post-Minkowskian  approximation for the scalar field equation}
Note that both source terms $\mu^{\alpha\beta}$ in~\eqref{eq:FEh} and $\mu_s$ in~\eqref{eq:finaleomphi} with~\eqref{eq:sourcescalar} and~\eqref{eq:Tmunu} depend on the fields $\varphi$ and $h^{\alpha\beta}$. These equations therefore have to be solved iteratively. In general during the inspiral, as the separation between the bodies is large, their mutual gravitational interaction is small and one can expand in weak gravitational fields. This is the so called Post-Minkowski (PM) expansion, traced by the factors of $G$. The field equations are in first instance solved iteratively order by order in $G$. On top of this in certain regions of spacetime additional approximations hold. In a region around the sources called the near zone we can expand in small velocities compared to the speed of light, the Post-Newtonian (PN) expansion. On the other hand far from the sources in the far field, one can expand in the large distance to the source resulting in a multipole expansion. These additional expansions on top of the PM expansion need to be meshed together to obtain the full solution for the fields throughout spacetime. 
 
To implement such an approximation scheme, we should bring the higher order terms in factors of $G$ to the RHS of the equation, as the baseline for solving the equation iteratively in the PM expansion in the coming sections. The squareroot of the metric determinant expands as follows
\be
\sqrt{-g} = 1-\frac{1}{2} h + \frac{1}{8} h^2 -\frac{1}{4}h^{\mu\nu}h_{\mu\nu}+\mathcal{O}(G^3),
\ee
with $h=h^{\mu\nu}\eta_{\mu\nu}$ and each factor of $h$ or $h_{\mu\nu}$ being $\mathcal{O}(G)$. The field equation for the scalar field up to second order in $G$ (the appropriate PM order to obtain the 1PN field solutions) is then given by
\be\label{eq:FEphiFS}
\bal
(\square_{\eta}-m_{\varphi}^2)\varphi &=\frac{4 \pi G}{c^4}(1+\frac{1}{2}h)\mu_s + h^{\mu\nu}\partial_{\mu}\partial_{\nu}\varphi -\frac{1}{2} m_{\varphi}^2 h \varphi+\mathcal{O}(G^3), \\
&\equiv\frac{4 \pi G}{c^4} \tilde{\mu}_s, 
\eal
\ee
where the effective source term $\tilde \mu_s$ to 2PM order can be read off from the RHS in the first line of~\eqref{eq:FEphiFS}.

\subsection{Integral solutions to the field equations}
With the above weak-field approximation for the scalar field equation to cast it as~\eqref{eq:FEphiFS}, both the scalar and the metric field equations~\eqref{eq:Boxh} involve as the highest-derivative operator the d'Alembertian of Minkowski space. Thus, we can write down the formal solutions as unevaluated integrals over the past lightcones $\mathcal{C}$ of the respective Greens functions $G(x,x')$ and the sources

\be\label{eq:formalsolhphi}
\begin{aligned}
    h^{\alpha \beta} &= \frac{4 G}{c^4} \int_{\mathcal{C}} G_{\square_{\eta}}(x,x')\mu^{\alpha\beta}(x')d^4x',\\
    \varphi &= -\frac{G}{c^4} \int_{\mathcal{C}} G_{\square_{\eta}-m_{\varphi}^2}(x,x') \tilde{\mu}_s(x') d^4 x',
\end{aligned}
\ee
with $x=(ct,\mathbf{x})$ and 
\begin{equation}
\bal
G_{\square_{\eta}}\left(ct-ct^{\prime}, \mathbf{x}-\mathbf{x'}\right)&=\frac{\delta\left(ct-ct^{\prime}-\left|\mathbf{x}-\mathbf{x'}\right|\right)}{\left|\mathbf{x}-\mathbf{x'}\right|},\\
G_{\square_{\eta}-m_{\varphi}^2}\left(ct-ct^{\prime}, \mathbf{x}-\mathbf{x'}\right)&=\frac{\delta\left(ct-ct^{\prime}-\left|\mathbf{x}-\mathbf{x'}\right|\right)}{\left|\mathbf{x}-\mathbf{x'}\right|}\\
&-\Theta\left(ct-ct^{\prime}-\left|\mathbf{x}-\mathbf{x'}\right|\right) \frac{m_{\varphi} J_1\left(m_{\varphi} \sqrt{\left(ct-ct^{\prime}\right)^2-\left|\mathbf{x}-\mathbf{x'}\right|^2}\right)}{\sqrt{\left(ct-ct^{\prime}\right)^2-\left|\mathbf{x}-\mathbf{x'}\right|^2}}.
\eal
\end{equation}
Here $\Theta$ is the Heaviside stepfunction and $J_1$ the Bessel function of the first kind~\cite{Alsing:2011er}. From the Greens function of the scalar field equation we can see that the scalar mass adds an additional contribution from inside the past lightcone related to the term proportional to the stepfunction.

\subsubsection{Overview of the direct integration method for computing the integrals} 
\begin{figure}[h!]
\centering
\begin{subfigure}{0.4\textwidth}
   \includegraphics[width=\linewidth]{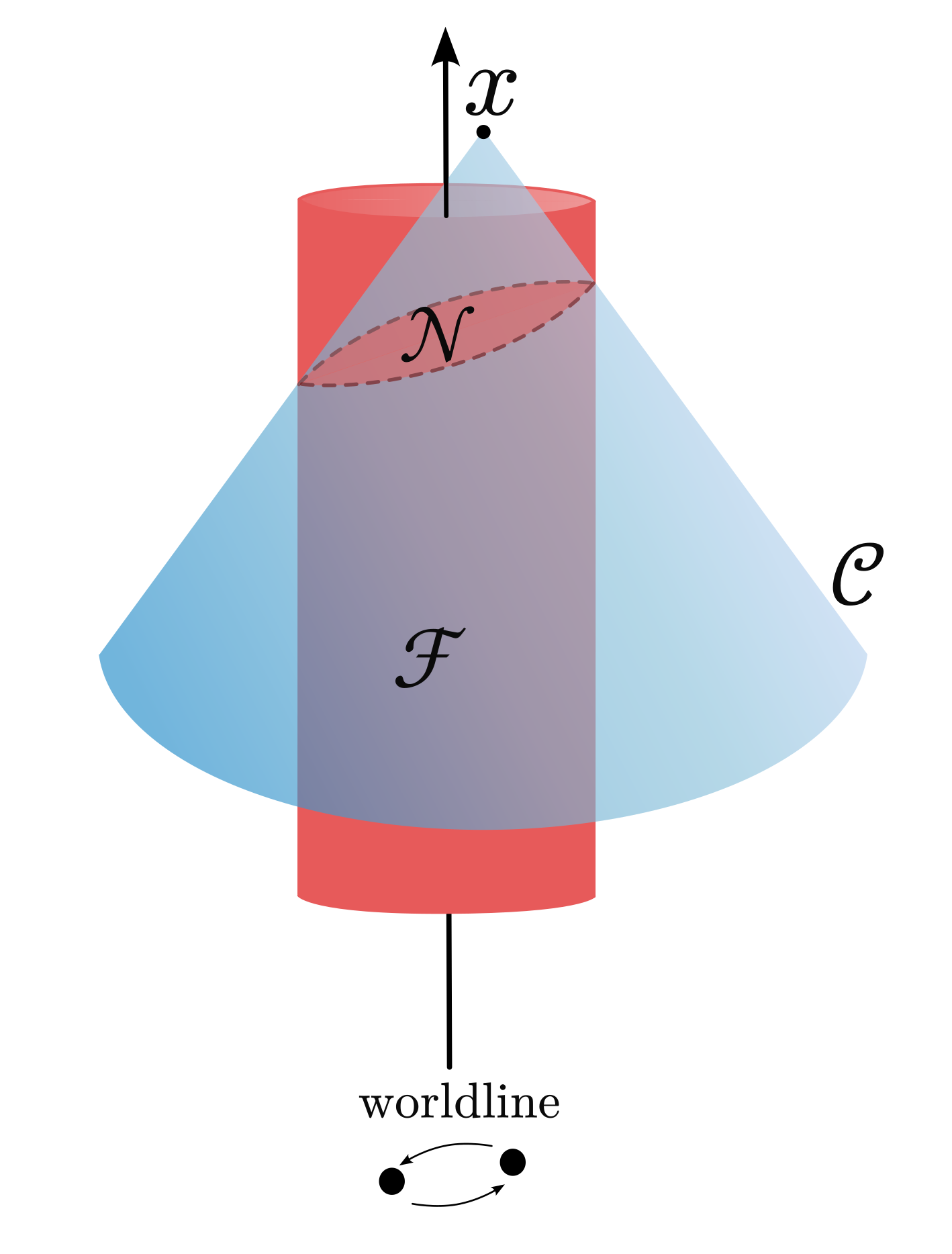}
       \centering
\end{subfigure}
\caption[]{Schematic figure showing the past lightcone (blue) from field point $x$ intersecting with the worldtube (red) of the scale of the binary system. The intersection hypersurface gives the near zone $\mathcal{N}$ and outside the far zone $\mathcal{F}$.}
\label{fig:DIRE}
\end{figure}

To compute the integrals we will use the DIRE approach (Direct Integration of the Relaxed Einstein equations) \cite{Pati:2000vt} for which the integration domain is split in the part of the lightcone $\mathcal{N}$ for which the cone intersects the so called near zone defined by radius $\mathcal{R}$ and far zone part of the lightcone $\mathcal{F}$ outside $\mathcal{R}$, see Fig.~\ref{fig:DIRE}. The near zone region is constructed as such that retardation effects are small and the PN expansion is valid, for the exact definition of the near zone, see \cite{Pati:2000vt}.\\
The full field solution then consists of the integration over the near zone and over the far zone part of the past lightcone. Additionally, the approximations one can do for these two integration domains depend on if the field point $x$ lies within the near or far zone. Essentially within DIRE there are four cases of integrals, near and far zone integrations for field point in the near zone and in the far zone. Each integrant can be expanded in a different way. For the purposes of this work we are interested in the equations of motion of the binary system and therefore we want to evaluate the fields near the sources in the near zone. We elaborate in the sections below how the integrands for the respective near zone and far zone parts of the past lightcone can be approximated. For an overview see \cite{Pati:2000vt}.\\
The final solution to the fields should not depend on the choice of near zone boundary $\mathcal{R}$ and therefore the boundary dependence of the near and far zone contributions have to cancel. This is shown in GR to hold for arbitrary PN order \cite{Pati:2000vt}. Here we explicitly check if this is still the case for the massive scalar field extension to leading order in $\mathcal{R}$ up to 1PN.

\section{Field solutions in the post-Newtonian approximation}\label{sec:fieldsol}

\subsection{Near zone integration}
In the near zone the PN approximation~\eqref{eq:PNapprox} holds, tracked by the orders in $c$. We derive the equations of motion (EoM) of the binary system up to first order in the PN expansion. To obtain the EoM up to 1PN, we require the metric $g^{\mu \nu}$ and $\varphi$ up to 1PN. Therefore we expand below the different components of the field equations in orders of $c$.

\subsubsection{PN expansion of the fields and source terms}
The leading order contributions of the point particle EM tensor \eqref{eq:Tmunu} are
\be
\begin{aligned}
& T^{00} \sim m_A(\varphi) c^2,\quad T^{0 j} \sim m_A(\varphi) v^j c,\quad T^{i j} \sim m_A(\varphi)v_A^i v^j \\
& \quad \Rightarrow T^{0 j} / T^{00} \sim \frac{v}{c},\quad T^{i j} / T^{00} \sim\left(\frac{v}{c}\right)^2,
\end{aligned}
\ee
or more directly: $T^{00} \sim \mathcal{O}\left(c^2\right), T^{0 j} \sim \mathcal{O}(c), T^{j k} \sim O(1)$ substituting in \eqref{eq:FEh} this results in the leading order contributions to $h^{\mu\nu}$ 

\be\label{eq:exph}
h^{00(0)} \sim \mathcal{O}\left(c^{-2}\right),\quad h^{0 j(0)}\sim \mathcal{O}\left(c^{-3}\right),\quad h^{j k^{(0)}}=\mathcal{O}\left(c^{-4}\right),
\ee
with (0) notation for the leading order PN contribution.\\

\noindent The metric contributions in terms of $h^{\mu \nu}$ are given via $\sqrt{-g} g^{\mu \nu}=\eta^{\mu\nu}+h^{\mu\nu}$ in powers of the PM expansion in $G\left(\eta^{\mu \nu} \sim \mathcal{O}(1), h^{\mu \nu} \sim \mathcal{O}(G)\right)$ 

\be
g_{\alpha \beta}=\eta_{\alpha \beta}+h_{\alpha \beta}-\frac{1}{2} h \eta_{\alpha \beta}+h_{\alpha \mu} h_\beta^\mu-\frac{1}{2} h h_{\alpha \beta}+\left(\frac{1}{8} h^2-\frac{1}{4} h^{\mu \nu} h_{\mu \nu} \eta_{\alpha \beta}\right)+\mathcal{O}\left(G^3\right).
\ee

\noindent The Newtonian metric in terms of $h^{\mu\nu}$ is 
\be
\begin{aligned}
g_{00} &= -1 + \frac{1}{2} h^{00(0)} + \mathcal{O}(c^{-4}),\\
g_{0i} &= 0  + \mathcal{O}(c^{-3}),\\
g_{ij} &= \delta_{ij} +  + \mathcal{O}(c^{-2}),
\end{aligned}
\ee
for the metric at 1PN we require the respective metric components to $1/c^2$ higher
\be\label{eq:gmunu1PN}
\begin{aligned}
& g_{00}^{(1)}=-1+\frac{1}{2}\left(h^{00(0)}+h^{00(1)}\right)-\frac{3}{8}\left(h^{00(0)}\right)^2+\frac{1}{2} h^i i^{(0)}+\mathcal{O}\left(c^{-6}\right), \\
& g_{0 j}^{(1)}=-h^{0 j(0)}+\mathcal{O}\left(c^{-5}\right), \\
& g_{j k}^{(1)}=\delta_{j k}\left(1+\frac{1}{2} h^{00(0)}\right)+\mathcal{O}\left(c^{-4}\right),\\
&\sqrt{-g} = 1+ \frac{1}{2} h^{00(0)} +\mathcal{O}\left(c^{-4}\right).
\end{aligned}
\ee
And for the scalar field
\be\label{eq:phi1PN}
\varphi=\varphi_{0}+ \delta\varphi^{(0)}+\delta \varphi^{(1)}+\mathcal{O}\left(c^{-4}\right),
\ee
with $\varphi_0$ the background value. 
The scalar dependent mass can be expanded as follows
\begin{eqnarray}
 M_{A}(\varphi)&=&m_{A}\left\{1+\alpha_A\delta \varphi^{(0)}_A+\left(\alpha_A \delta\varphi^{(1)}_A\right.\right.\left.\left.+\frac{1}{2}\left[\left(\alpha_A\right)^2+\beta_A\right]\delta {\varphi^{(0)}_A}^2\right)\right\} +\mathcal{O}(c^{-6})\,,\qquad 
\end{eqnarray}
with coefficients
\begin{equation}
    \left.\alpha_{A}=\frac{d\, \log m_{A}(\varphi)}{d\varphi}\right|_{\varphi=\varphi_{0}},
    \quad
    \left.\beta_{A}=\frac{d\, \alpha_{A}(\varphi)}{d\varphi}\right|_{\varphi=\varphi_{0}}.\quad 
\end{equation}\\
The scalar dependent mass and $\alpha_A$, $\beta_A$ are at this stage just coefficients. One needs to match the field solutions to the full compact object solution~\cite{vanGemeren:2024bzf} to link the coefficients to physical quantities. 

\noindent The PN expanded source terms \eqref{eq:sourcetensor}, \eqref{eq:sourcescalar} of field equations are given by
\be\label{eq:PNsources}
\begin{aligned} \mu^{00}= & \sum_A m_A c^2\left[1+\frac{v_A^2}{2 c^2}+\frac{3}{4} h^{00(0)}+\alpha_A \delta \varphi^{(0)}\right] \delta^3(x-x_A(t)) \\ & -\frac{c^4}{16 \pi G}\left[\frac{7}{8}\left(\nabla h^{00(0)}\right)^2-2\left(\nabla \delta \varphi^{(0)}\right)^2\right] \\ & +2\frac{ \alpha f'\left(\varphi_0\right) c^4}{16 \pi G} \partial_{i j}\left(\delta \varphi^{(0)}\right) \partial_{i j}\left(h^{00(0)}\right)+\frac{c^4}{16 \pi G} m_{\varphi}^2\left(\delta \varphi^{(0)}\right)^2+\mathcal{O}\left(c^{-2}\right), \\ \mu^{0 j}= & \sum_A m_A c v_A^j \delta^3(x-x_A(t))+\mathcal{O}\left(c^{-1}\right), \\ \mu^i i= & \sum_A m_A v_A^2 \delta^3(x-x_A(t))+\frac{c^4}{16 \pi G}\left(-\frac{1}{8}\left(\nabla h^{00(0)}\right)^2-2\left(\nabla \delta \varphi^{(0)}\right)^2\right) \\ & -2\frac{\alpha f'\left(\varphi_0\right) c^4}{16 \pi G}\left( \nabla^2 \delta \varphi^{(0)} \nabla^2 h^{00(0)}\right)+\frac{c^4}{16 \pi G} m_{\varphi}^2\left(\delta \varphi^{(0)}\right)^2+\mathcal{O}\left(c^{-2}\right), \\ \mu_s= & \sum_A m_A c^2 \alpha_A \delta^3(x-x_A(t))\left(1-\frac{v_A^2}{2 c^2}-\frac{3}{4} h^{00(0)}+\left(\alpha_A+\frac{\beta_A}{\alpha_A}\right) \delta \varphi^{(0)}\right) \\ & -\frac{\alpha f^{\prime}\left(\varphi_0\right) c^4}{32 \pi G}\left(\left(\partial_{k l} h^{00(0)}\right)^2-\left(\nabla^2 h^{00(0)}\right)^2\right)+\mathcal{O}\left(c^{-2}\right).\end{aligned}
\ee\\
These expansions will be substituted in the field equations~\eqref{eq:FEh},\eqref{eq:FEphiFS} and solved per order in $1/c$ till $\mathcal{O}(c^{-4})$ which corresponds to the 1PN order. 

\subsubsection{PN expansion of the field integrals}

With these expansions at hand we can already see that the $h^{\mu\nu}\partial_{\mu}\partial_{\nu}\varphi$ term in~\eqref{eq:FEphiFS} will be of order $\mathcal{O}(c^{-6})$ so would not contribute to the 1PN contributions. 

The integral form of the scalar field solution simplifies even more when considering how the Greens function reduces in the PN approximation. This is most conveniently shown in momentum space. The Klein-Gordon operator in momentum space is given by
\be
G_{\square_{\eta}-m_{\varphi}^2}\left(t-t^{\prime}, \mathbf{x}-\mathbf{x'}\right)= \int \frac{d\omega}{2 \pi}\int \frac{d^3 k}{(2\pi)^3} \frac{e^{- i \omega (t-t')}e^{i k(x-x')}}{-k^2+\frac{\omega^2}{c^2}-\mphi^2}
\ee
In the near zone at the orbital scale, the orbital frequency and wavenumber scale as $\omega \sim \frac{v}{c|x-x'|}$ and $|k|\sim \frac{1}{|x-x'|}$. For small velocities this implies the hierarchy $\frac{\omega}{c}\ll|k|$ and expanding the denominator in small $\frac{\omega}{c |k|}$ to 1PN results in 
\be
\bal
G_{\square_{\eta}-m_{\varphi}^2}\left(t-t^{\prime}, \mathbf{x}-\mathbf{x'}\right)&= -\int \frac{d\omega}{2 \pi} e^{- i \omega (t-t')} \int \frac{d^3 k}{(2\pi)^3} \frac{e^{i k(x-x')}}{k^2+\mphi^2}\\
&-\int \frac{d\omega}{2 \pi}\frac{\omega^2}{c^2} e^{- i \omega (t-t')} \int \frac{d^3 k}{(2\pi)^3} \frac{e^{i k(x-x')}}{(k^2+\mphi^2)^2}\\
&=-\delta(t-t') \frac{e^{-\mphi |x-x'|}}{|x-x'|} -\frac{1}{c^2} \partial_t^2 \delta(t-t') \frac{e^{-\mphi |x-x'|}}{2 \mphi}.
\eal
\ee
Then the solution to the scalar field \eqref{eq:FEphiFS} becomes
\be
\bal
 \varphi  &= \frac{G}{c^4}( - \frac{e^{-\mphi |x-x'|}}{|x-x'|}\tilde{\mu}_s(t, x')  d^3 x' -\int \frac{1}{c^2}\delta(t-t') \partial_t^2 \bigg(\frac{e^{-\mphi |x-x'|}}{2 \mphi} \tilde{\mu}_s(t',x')\bigg)  d t'd^3 x'\\
 &=-\frac{G}{c^4}( \int \frac{e^{-\mphi |x-x'|}}{|x-x'|}\tilde{\mu}_s(t, x')  d^3 x' +\frac{1}{c^2}\int \partial_t^2 \bigg(\frac{e^{-\mphi |x-x'|}}{2 \mphi} \tilde{\mu}_s(t,x')\bigg) d^3 x'.
 \eal 
 \ee
 As the second derivative term already carries the order $c^{-2}$ only the $c^{2}$ order contribution from the source term \eqref{eq:PNsources} contributes to 1PN, which is a constant term. Up to this order we can therefore write the scalar field as
 \be\label{eq:FIphiPN}
\bal
 \varphi  &=-\frac{G}{c^4}( \int \frac{e^{-\mphi |x-x'|}}{|x-x'|}\tilde{\mu}_s(t, x')  d^3 x' +\frac{1}{c^2} \int \tilde{\mu}_s^{(0)}\partial_t^2 \bigg(\frac{e^{-\mphi |x-x'|}}{2 \mphi}\bigg) d^3 x'.
 \eal 
 \ee

\noindent For the $h^{\alpha\beta}$ fields we can do the following. After integrating out the time dependence, the field integrals in the PN approximation are given by
\be\label{eq:FIh00PN}
\begin{aligned}
    h^{\alpha \beta} &= \frac{4 G}{c^4} \int_{\mathcal{M}} \frac{\mu^{\alpha\beta}(t-|x-x'|,x')}{|x-x'|} d^3x'.\\
\end{aligned}
\ee
With $\mathcal{M}$ the constant time hypersurface at $t$ over the near zone up to boundary $\mathcal{R}$.
In this case both the field point $x$ as the integration variable $x'$ lie in the near zone. Therefore $|x-x'|$ is small and we can Taylor expand the time dependence of the source functions 
\be
\mu^{\alpha\beta}(t-|x-x'|,x') = \mu^{\alpha\beta}(t,x')-\frac{1}{c} \frac{\partial \mu^{\alpha\beta}(t,x')}{\partial t} |x-x'| + \frac{1}{2 c^2} \frac{\partial^2 \mu^{\alpha\beta}(t,x')}{\partial t^2} |x-x'|^2 + \ldots
\ee
For the 1PN fields we're interested in the corrections up to $\mathcal{O}(c^{-4})$, as the source~\eqref{eq:PNsources} start at $\mathcal{O}(c^{2})$ for $h^{00}$, at $\mathcal{O}(c^{1})$ for $h^{0j}$, at $\mathcal{O}(c^{0})$ for $h^{ij}$, we keep up to second, first, zeroth order in the time derivatives respectively. Executing the expansion in $|x-x'|$ keeping the terms up the the relevant order in $c$ we obtain

\be\label{eq:expintegralsh}
\begin{aligned}
    h^{00}_{\mathcal{N}} &= \frac{4 G}{c^4}\left(\int_{\mathcal{M}}\frac{\mu^{00}}{|x-x'|}d^3x' -\frac{1}{c}\frac{\partial}{\partial t}\int_{\mathcal{M}} \mu^{00}d^3x' +\frac{1}{2 c^2}\frac{\partial^2}{\partial t^2}\int_{\mathcal{M}} \mu^{00} |x-x'| d^3x'\right),\\
     h^{0j}_{\mathcal{N}} &= \frac{4 G}{c^4}\left(\int_{\mathcal{M}}\frac{\mu^{0j}}{|x-x'|}d^3x' -\frac{1}{c}\frac{\partial}{\partial t}\int_{\mathcal{M}} \mu^{0j}d^3x'\right),\\
      h^{ij}_{\mathcal{N}} &= \frac{4 G}{c^4}\left(\int_{\mathcal{M}}\frac{\mu^{ij}}{|x-x'|}d^3x'\right).
\end{aligned}
\ee
One can rewrite the linear time derivative contributions to surface integrals using the conservation equation $\partial_{\beta}\mu^{\alpha\beta}=0$. It can be shown their contribution only enters at high PN order, we will discard these contributions from now on, see section 7.1.2 in \cite{Poisson:2014}. 
For the scalar field, the time dependence was already expanded in PN orders by expanding the Greens function, so after taking the time derivatives, we only expand the exponentials
 \be\label{eq:expintegralphi}
\bal
 \varphi_{\mathcal{N}}  &=-\frac{G}{c^4}( \int_{\mathcal{M}} \sum_{\ell=0}^{\infty} \frac{(-1)^{\ell}}{\ell!} \frac{\partial^L}{\partial |x-x'| ^L}\left( \frac{e^{-\mphi |x-x'|}}{|x-x'|}\right)\bigg|_{|x-x'|\xrightarrow{}0}\tilde{\mu}_s(t, x')  d^3 x'\\
 &+\frac{1}{c^2} \int_{\mathcal{M}} \tilde{\mu}_s^{(0)}\partial_t^2 \sum_{\ell=0}^{\infty} \frac{(-1)^{\ell}}{\ell!} \frac{\partial^L}{\partial |x-x'| ^L}\bigg(\frac{e^{-\mphi |x-x'|}}{2 \mphi}\bigg)\bigg|_{|x-x'|\xrightarrow{}0} d^3 x'.
 \eal 
 \ee

\noindent We're now prepared to substitute the PN expansion of the fields \eqref{eq:exph}, \eqref{eq:phi1PN} and source terms \eqref{eq:PNsources} in the integral solutions \eqref{eq:expintegralsh}, \eqref{eq:expintegralphi} above and collect the terms per order in the expansion. To explicitly evaluate the field integrals we will cut of the Taylor expansion to $\ell=2$ for the scalar field equation from now on. 

\subsubsection{0 PN field solutions}

From substituting the field expansion \eqref{eq:phi1PN} in \eqref{eq:expintegralphi} one can immediately see that at order $\mathcal{O}(c^{0})$ the field has no source. Therefore the background scalar field $\varphi_0=0$, compared to the massless scalar field solution where the background field is a finite constant~\cite{Julie:2019sab}. \\

\noindent At order $\mathcal{O}(c^{-2})$ only $\varphi$ and $h^{00}$ have contributions for which the integration can be straightforwardly done as the source terms are proportional to the dirac delta. 
\be
\bal
h^{00(0)}_{\mathcal{N}}=& \frac{4 G m_A}{c^2 |x-x_A(t)|} + \AB,\\
\delta\varphi^{(0)}_{\mathcal{N}}=& -\frac{G m_A \alpha_A}{c^2} \left(\frac{1}{|x-x_A(t)|} - \mphi +\mphi^2\frac{|x-x_A(t)|}{2}\right)+ \AB.
\eal
\ee

\noindent We will see in section \ref{sec:BinaryDyn} we require the fields evaluated at the location of the bodies to obtain the two body dynamics. Hence we would like to set $x=x_B$ (the reversed case $x=x_A$ can be obtained in the end by reversing the body labels.) This results in a more subtle approach to evaluate the field integrals. If we look again at $h^{00(0)}$ for $x=x_B$ we have
\be
\bal
h^{00(0)}_{\mathcal{N}}=\frac{4 G}{c^4} \left(\int \frac{m_A c^2}{|x_B-x'|}\delta^3(x'-x_A(t))d^3 x' + \int\frac{m_B c^2}{|x_B-x'|}\delta^3(x'-x_B(t))d^3 x'\right),
\eal
\ee
where the first integral can be evaluated straightforwardly, the integration over $\frac{1}{|x_B-x'|}\delta^3(x'-x_B(t))$ is ill defined. This is a consequence of the point particle approximation which breaks down at the location of the bodies. One can still make sense of these contributions using an appropriate regularization scheme. To resolve this issue to high order in the post-Newtonian expansions in GR the Hadamard regularization is used \cite{Hadamard1932,Schwartz1978,Bel:1981be, Damour:1981bh,Schaefer:1985vxb,Blanchet:1995fg,Jaranowski:1997ky,Blanchet:1998vx}. For defining this regularization scheme we mostly follow \cite{Blanchet:2000nu}. 
We define the functions $F(x)$ on $\mathbb{R}^3$ that we integrate over to be smooth on the whole domain except from the points $x_A$, $x_B$. One can expand the function $F(x)$ around these points via a Laurent series
\be\label{eq:LaurentF}
F(\mathbf{x})=\sum_{n=0}^{n_{max}} r_A^{-n} f_{A,n}\left(\mathbf{n}_A\right) + \mathcal{O}(r_A),
\ee
with $r_A=|x-x_A|$ and $n_A = \frac{\mathbf{r_A}}{r_A}$, in the same way we can write the expansion around $r_B$. This expansion diverges at the point $x=x_A$ and is regularized by defining the `Hadamard partie finie' (PF) at the singular point
\be\label{eq:PFdef}
(F)_A=\int \frac{d \Omega_A}{4 \pi}\,f_{A,0}\left(\mathbf{n}_A\right),
\ee
with $\Omega_A(\mathbf{n}_A)$ the element of the solid angle in the direction of $\mathbf{n}_A$. In other words the PF is the angular average of the zeroth order coefficient of the Laurent series. We can use the PF to interpret the integration of F times a dirac delta by defining
\be
F(x')\delta^3(x'-x_B(t)) = (F)_B\,\delta^3(x'-x_B(t)), 
\ee
thus $\int F(x')\delta^3(x'-x_B(t)) d^3x' = (F)_B$. Now in the case for the integration in $h^{00(0)}$ evaluated at $x=x_B$ we have $F(x') = \frac{m_B c^2}{|x_B-x'|}$ which has $(F)_B=0$ as it has no term in it's Laurent series proportional to $r_B^0$. Thus the contribution of this integral vanishes. \\

\noindent The field solutions at 0PN order for $h^{00(0)}$ and $\delta \varphi^{(0)}$ evaluated at $x=x_B$ are then given by
\be\label{eq:sol0PNh00phi}
\bal
h^{00(0)}_{\mathcal{N}}(x_B)=& \frac{4 G m_A}{c^2 r_{AB}},\\
\delta\varphi^{(0)}_{\mathcal{N}}(x_B)=& -\frac{G m_A \alpha_A}{c^2} \left(\frac{1}{r_{AB}} - \mphi +\mphi^2\frac{r_{AB}}{2}\right)+ \frac{G m_B \alpha_B}{c^2}\mphi.
\eal
\ee

\noindent In the same fashion we obtain at order $\mathcal{O}(c^{-3})$ 
\be\label{eq:sol05PNh0j}
h^{0j(0)}_{\mathcal{N}}(x_B) = \frac{4 G m_A v_A^j}{c^3 r_{AB}},
\ee
and at order $\mathcal{O}(c^{-4})$
\be\label{eq:sol1PNhii}
{h^{i(0)}_i}_{\mathcal{N}}(x_B) = \frac{4 G m_A v_A^2}{c^4 r_{AB}}.
\ee

\subsubsection{1 PN field solutions}\label{sec:1PNNZfields}
At order $\mathcal{O}(c^{-4})$ we substitute the 0PN solutions as functions of $x$ in the sources \eqref{eq:PNsources}. To rewrite terms $(\nabla h^{00(0)})^2$ and $(\nabla \delta\varphi^{(0)})^2$ in \eqref{eq:PNsources} we can use $(\nabla h^{00(0)})^2 = \frac{1}{2}\nabla^2{h^{00(0)}}^2 - h^{00(0)}\nabla^2h^{00(0)}$, same for the scalar field. Furthermore we can replace $\nabla^2 h^{00(0)}$ and $\nabla^2\delta\varphi^{(0)}$ by recognizing that the zeroth order integrals $h^{00(0)}=\frac{4 G }{c^4}\int \frac{\mu^{00(0)}}{|x-x'|}d^3x'$ and $\delta\varphi^{(0)}=\frac{-G}{c^4}\int \frac{e^{-\mphi |x-x'|}}{|x-x'|}\tilde{\mu}_{s}^{(0)}$ are equivalent to $\nabla^2 h^{00(0)} = -\frac{16 \pi G}{c^4} \mu^{00(0)}$ and $(\nabla^2-\mphi^2)\delta\varphi^{(0)}=\frac{4\pi G}{c^4}\tilde{\mu}_{s}^{(0)}$. \\

\noindent The 1PN integrals of the fields consist of two types. The contributions with compact support, which again evaluate straightforwardly over the dirac deltas, and purely field dependent term. The latter require a more subtle treatment. We'll explain explicitly how we evaluate the $\alpha$ dependent term in the $h^{00(1)}$ field. All the other purely field dependent contributions can be treated in the same way. 

\be
\bal
h^{00(1)}_{\mathcal{N}} \propto& \frac{4 G}{c^4} \frac{2 \alpha f'(\varphi_0) c^4}{16 \pi G} \int_{\mathcal{M}} \frac{1}{|x-x'|}\frac{\partial^2}{\partial x'^i \partial x'^j}\delta\varphi^{(0)}\frac{\partial^2}{\partial x'^i \partial x'^j}h^{00(0)}d^3x' \\
\propto&  -\sum_{A,B} \frac{8 \alpha f'(\varphi_0) c^4}{ 4 \pi} \frac{G^2 m_A m_B \alpha_A}{c^4} \int_{\mathcal{M}} \frac{1}{|x-x'|} \frac{\partial^2}{\partial x'^i \partial x'^j}\left(\frac{1}{|x'-x_A(t)|} - \mphi +\mphi^2\frac{|x'-x_A(t)|}{2}\right)\\
&\frac{\partial^2}{\partial x'^i \partial x'^j}\bigg(\frac{1}{|x'-x_B(t)|}\bigg)d^3x'.
\eal
\ee
Here the sum notation stands for the sum over the different permutations of A, B. In the massless scalar field theory one can make use of the following. As the derivatives to $x$ are always evaluated on the lengths $|x-x_{A/B}|$ one can change the derivation to $\frac{\partial}{\partial x^i} |x-x_A| = -\frac{\partial}{\partial x_A^i}|x-x_A|$. 
\be\label{eq:GBintml}
h^{00(1)}_{\mathcal{N},\mphi=0} \propto  -\sum_{A,B} \frac{8 \alpha f'(\varphi_0) c^4}{ 4 \pi} \frac{G^2 m_A m_B \alpha_A}{c^4} \frac{\partial^2}{\partial x_A^i \partial x_A^j} \frac{\partial^2}{\partial x_B^i \partial x_B^j} \int_{\mathcal{M}} \frac{1}{|x-x'|} \frac{1}{|x'-x_A| |x'-x_B|}.
\ee
This integral has the closed form solution \cite{Blanchet:2013haa,Poisson:2014}
\be\label{eq:mlGBlogsol}
\int_{\mathcal{M}} \frac{1}{|x-x'|} \frac{1}{|x'-x_A| |x'-x_B|} = 4\pi \left(1-\log\left(\frac{|x-x_A|+|x-x_B|+|x_A-x_B|}{2 \mathcal{R}}\right)\right).
\ee
Unfortunately for other (inverse) powers of the lengths in the integrant, as we have for example at quadratic order in the scalar mass, we could not find a closed form solution. Additionally for higher powers of inverse lengths these integrals have to be regularized at the locations of the bodies $x_A$, $x_B$. Therefore we set about a general way to explicitly evaluate the integrals. \\

\noindent If we again look at the higher curvature massless scalar field term in $h^{00(1)}$ but leaving the derivatives inside the integration
\be\label{eq:GBint}
\bal
\int_{\mathcal{M}} \frac{1}{|x-x'|} \frac{\partial^2}{\partial x'^i \partial x'^j}\frac{1}{|x'-x_A|} \frac{\partial^2}{\partial x'^i \partial x'^j}\frac{1}{|x'-x_B|
}d^3x'&= \int_{\mathcal{M}} \frac{1}{|x-x'|}\bigg(-\frac{1}{|x'-x_A|^3 |x'-x_B|^3}\\
&+ \frac{9(n'_A\cdot n'_B)^2}{|x'-x_A|^3 |x'-x_B|^3}\bigg)d^3x',
\eal
\ee
with $n^{'i}_{A/B} = \frac{(x'-x_{A/B})^{i}}{|x'-x_{A/B}|}$. For these types of integrals we integrate over source points $x'$ inside the near zone and the field evaluates at field point $x$. For all our following purposes we are interested in evaluating the fields at one of the bodies, so we can set $x$ to $x_A$ or $x_B$. There actually is a subtlety to considering this limit of the field point which we will come back to at the end of this section. For now we focus on evaluating integrals of the type 
\be\label{eq:Fintgen}
\bal
\int_{\mathcal{M}} \frac{1}{|x'-x_A|^{\alpha} |x'-x_B|^{\beta}}d^3x'.
\eal
\ee

\noindent These type of functions are smooth everywhere except at the location of the bodies where they become singular so we again need to resort to the Hadamard regularization to evaluate these integrals. For the following discussion we again mostly follow \cite{Blanchet:2000nu}.\\ 

\noindent If one wishes to integrate $F(x)$ over the domain containing the singular points, one can define the integral as the part of this integral over $F$ over the domain except for the singular points plus the PF of the function at these singular points. Formally one defines two spherical balls $\mathcal{B}_A(s)$, $\mathcal{B}_B(s)$ of radius $s$ around the singular points, assuming $s$ is small so the balls do not intersect. The diverging parts of the integral involve, after integration, polynomial terms of $1/s$ for powers smaller than $n<3$ in the Laurent series and a logarithmic term in $s$ for $n=1$, with $n$ being $\alpha$ or $\beta$ \eqref{eq:Fintgen}. Because of the logarithmic term in $s$ an ambiguity arises due to the freedom to rescale and can be seen as the freedom of choice of unit system for the length of $s$. One can capture this ambiguity by introducing the lengths $s_A$, $s_B$ to also remove the dimension from the logarithms. All in all we have for the PF integral
\be\label{eq:PFInt}
\begin{aligned}
\mathrm{Pf}_{s_A, s_B} \int d^3 \mathbf{x} F=&\lim _{s \rightarrow 0}\bigg\{\int_{\mathbb{R}^3 \backslash \mathcal{B}_A(s) \cup \mathcal{B}_B(s)} d^3 \mathbf{x} F+\bigg(\sum_{a+3<0} \frac{s^{a+3}}{a+3} \int d \Omega_A f_{A,a}+\ln \left(\frac{s}{s_A}\right) \int d \Omega_A f_{A,-3} \\
&+\AB\bigg) \bigg\}.
\end{aligned}
\ee
\\
In practice one evaluates the integral over $F$ by changing to spherical coordinates to evaluate the angular and radial integration. It's convenient to employ a change of variables from $x'$ to one of the lengths $r_A$ or $r_B$. In section IV part B of \cite{Blanchet:2000nu} the authors discuss an approach to change the integration variable to $r_A$ and regularize $F$ by subtracting the terms diverging at $r_B$. The procedure in this reference is done for an infinite integration domain. We walk through the same steps in Appendix~\ref{sec:appendixA} but this time for a finite domain restricted to the near zone boundary $\mathcal{R}$. We give the final result here 

\begin{equation}\label{eq:PFintexpl}
\bal
\operatorname{Pf} \int d^3 \mathbf{x} F
&=\lim _{s \rightarrow 0}\left\{\int d\Omega_A \int_{s}^{\mathcal{R}} d r_A \tilde{F}_B r_A^2 - \int d\Omega_A \tilde{F}_B \mathcal{R}^2 \mathbf{x}_A\cdot \mathbf{N_{r_A}}\right.\\
&\left. +\sum_{b+3<0} \frac{s^{b+3}}{b+3} \int d \Omega_A f_A+\ln \left(\frac{s}{s_A}\right) \int d \Omega_A f_A\right.\\
&\left. +\sum_{b+3<0}\frac{\mathcal{R}^{b+3}}{b+3} \int d\Omega_B f_{B,b}+\ln(\frac{\mathcal{R}}{s_B})\int d\Omega_B f_B^{-3} -\sum_{b+3\leq0}\int d\Omega_B\, r_B^{b} f_{B,b} \mathcal{R}^2 \mathbf{x_B}\cdot \mathbf{N_{rB}}\right\},
\eal
\end{equation}
with $
\tilde{F}_B = F-\sum_{b+3\leq 0} r_B^b f_{B,b}$
and $\boldsymbol{N_{r_{B/A}}}:=\boldsymbol{r_{B/A}} / r_{B/A}$. By doing the change of variables we approximated the integration domain up to order $O\left(x_B^2 / \mathcal{R}\right)$. Compared to the integration in \cite{Blanchet:2000nu} we obtain extra boundary related terms and we don't require to impose a finite part procedure to regularize the boundary at infinity.

We explicitly work out the computation of the PF integral of the fiducial example
\be
F(x) = \frac{1}{r_A^4 r_B^3}.
\ee
To obtain $\tilde{F}_B$ we require the Laurent expansion of F around $x_B$, in which $r_A$ is expanded around $r_{AB}$ 
\be
r_A^{-4}=\sum_{l \geq 0} \frac{(-)^l}{l!} r_B^l n_B^L \partial_L r_{AB}^{-4},
\ee
and we find for the Laurent series
\be
F(x)|_{x\xrightarrow{}x_B} =\frac{1}{r_{AB}^4 r_B^3} + \mathcal{O}(r_B^{-2}),
\ee
thus 
\be
\tilde{F}_{B} = \frac{1}{r_A^4 r_B^3} - \frac{1}{r_{AB}^4 r_B^3}.
\ee
Then the first integral of \eqref{eq:PFintexpl}, changing the integration variable to $r_A$ and expressing $r_B$ in terms of $r_A$ and $r_{AB}$; $r_{B} = \sqrt{r_A^2 + 2 r_{AB} r_A \cos(\theta) + r_{AB}^2}$, the angular integration has two outcomes depending on if $r_{AB}$ is smaller or larger than $r_A$, so the integration becomes
\be
\bal
\int d\Omega_A \int_{s}^{\mathcal{R}} d r_A \tilde{F}_B r_A^2 &=
\int d\Omega_A \int_{s}^{r_{AB}} d r_A \tilde{F}_B r_A^2 + 
\int d\Omega_A \int_{r_{AB}}^{\mathcal{R}} d r_A \tilde{F}_B r_A^2\\
&= 4\pi (-\frac{s}{r_{AB}^5} - \frac{1}{2 r_{AB}^4} +\log(\frac{r_{AB}}{\mathcal{R}}) + \frac{1}{s r_{AB}^3} + \frac{1}{2 \mathcal{R}^2 r_{AB}^2}).
\eal
\ee
In a similar way we expand $F$ around $x_B$ and change of variable to $r_B$ for the integrals in the final line of \eqref{eq:PFintexpl}. Finally the other integrations give
\be
\int d\Omega_A \tilde{F}_B \mathcal{R}^2 \mathbf{x}_A\cdot \mathbf{N_{r_A}} = 4\pi (\frac{|x_A|}{\mathcal{R}^2 r_{AB}^3}+\frac{|x_A|}{\mathcal{R}^4 r_{AB}}).
\ee

\be
\sum_{b+3<0} \frac{s^{b+3}}{b+3} \int d \Omega_A f_A+\ln \left(\frac{s}{s_A}\right) \int d \Omega_A f_A = -4\pi \frac{1}{s r_{AB}^3}.
\ee

\be
\sum_{b+3<0}\frac{\mathcal{R}^{b+3}}{b+3} \int d\Omega_B f_{B,b}+\ln(\frac{\mathcal{R}}{s_B})\int d\Omega_B f_B^{-3} = 4\pi \frac{\log(\frac{\mathcal{R}}{s_B})}{r_{AB}^4}.
\ee

\be
\sum_{b+3\leq0}\int d\Omega_B\, r_B^{b} f_{B,b} \mathcal{R}^2 \mathbf{x_B}\cdot \mathbf{N_{rB}} =0.
\ee
Substituting all the contributions back in \eqref{eq:PFintexpl} we obtain after taking the limit of $s\xrightarrow{}0$
\be
\operatorname{Pf} \int d^3 \mathbf{x} F = -4\pi \bigg(\frac{1}{2 r_{AB}^4} - \frac{\log(\frac{r_{AB}}{s_B})}{r_{AB}^4} -\frac{1}{2 \mathcal{R}^2 r_{AB}^2}  + \frac{|x_A|}{\mathcal{R}^2 r_{AB}^3} + \frac{|x_A|}{\mathcal{R
}^4 r_{AB}}\bigg). 
\ee

\noindent Doing this computation again but for the more complicated expression \eqref{eq:GBint} using $n_A \cdot n_B=\frac{r_A^2+r_B^2-r_{AB}^2}{2 r_A r_B}$ and expanding the final answer to quadratic order in $1/\mathcal{R}$ we obtain
\be
\operatorname{Pf}\int_{\mathcal{M}} (-\frac{3}{|x'-x_A|^3 |x'-x_B|^4} + \frac{9(n'_A\cdot n'_B)^2}{|x'-x_A| |x'-x_B|^4})d^3x' = -4\pi\frac{1}{2 r_{AB}^4} + \mathcal{O}(\frac{1}{\mathcal{R}^3}).
\ee
As a consistency check we can take the derivatives to $x_A$ and $x_B$ of the closed form solution we already obtained \eqref{eq:mlGBlogsol} and taking the limit $x\xrightarrow{}x_B$
\be
\frac{\partial^2}{\partial x_A^i \partial x_A^j} \frac{\partial^2}{\partial x_B^i \partial x_B^j}\left( 4\pi \left(1-\log\left(\frac{|x-x_A|+|x-x_B|+|x_A-x_B|}{2 \mathcal{R}}\right)\right)\right) = \lim_{s\xrightarrow{}0}[-4\pi (\frac{1}{2 r_{AB}^4} +\frac{1}{2 r_{AB}^3 s}) ]
\ee
for which the finite term agrees with our result above and with the literature on massless GB theory \cite{Julie:2019sabErr}.\\

\noindent Lastly we come back to the note we made at the beginning by specifying the field point $x$ to exactly be on the locations of the body as this is where we need the fields evaluated for the binary dynamics computations. However for the full field solutions one would have to keep $x$ generic. Also treating properly the limit of $x\xrightarrow{}x_{A/B}$. Defining the Poisson integral of $F$
\be
P(x) = \frac{1}{4\pi} \int d^3x' \frac{F(x')}{|x-x'|}.
\ee
The limit $P$ when the field point approaches the body locations is not continuous. Handling this limit in the correct manner~\cite{Blanchet:2000nu}, again defining the partie finie for $P$ the integral evaluated as $x\xrightarrow{}x_A$ is given by
\be
(P)_A=-\frac{1}{4 \pi} \mathrm{Pf}_{s_A, s_B} \int \frac{d^3 \mathbf{x'}}{|x'-x_A|} F(\mathbf{x'})+\left[\ln \left(\frac{|x-x_A|}{s_A}\right)-1\right]\left(|x'-x_A|^2 F\right)_A.
\ee
So the partie finie of the Poisson integral in the limit to $x_A$ is the partie finie of $x=x_A$ (this is what we obtained in the derivations above) with a correction term. The constant $s_A$ will cancel in the final expression~\cite{Blanchet:2000nu} and the partie finie will depend on the constants $\ln(s_B)$ and the infinite constant $\ln(|x-x_A|)$. It has been shown~\cite{Blanchet:2000nv, Blanchet:2000ub} that the infinite constants  and the same for for the other particle B, are unphysical as they can be removed by a coordinate transformation. Within Hadamard regularization the $s_B$ and $s_A$ constants for the other particle should be kept. These ambiguous parameters can be removed by a dimensional regularized treatment of the field integrals. This is shown in~\cite{Blanchet:2013haa}. In this work we won't do this derivation explicitly but assume these constants can be removed by adding the correction from dimensional regularization and neglect them in the rest of this work. Furthermore as for our purposes we are interested in the fields exactly evaluated at the body locations, the computation of the partie finie evaluated with the field point at the body locations suffices. \\

\noindent With this approach we evaluate the field dependent source terms in the 1PN field integrals. After integration we use for the time derivative terms in \eqref{eq:expintegralsh}, \eqref{eq:expintegralphi}; $\p_t^2|x-x_A| = \p_t(-\frac{(x-x_A)^i {v_A}_i}{|x-x_A|}) = -n_A^i {a_A}_i + \frac{v_A^2}{|x-x_A|} - \frac{(n_A\cdot v_A)^2}{|x-x_A|} $ with $v_A^i = \p_t x_A^i$ and $n_A^i = \frac{(x-x_A)^i}{|x-x_A|}$ and similar for $x_B$. Below we give the total solution of the metric and scalar fields up to 1PN to leading order in $\mathcal{R}$ evaluated at $x=x_B$. The fields for $x=x_A$ can be obtained by $\AB$ and we defined $r_{AB} = |x_A-x_B|$. To keep the length of the expressions manageable, we gave the explicit terms of the coefficients in Appendix~\ref{sec:AppB}.

\be\label{eq:NZsolh00}
\bal
h^{00(1)}_{\mathcal{N}}(x_B) &= \frac{G m_A}{c^4 r_{AB}} \left(\mathcal{A}_{h^{00}} + \frac{G m_A}{r_{AB}}\left( \mathcal{B}_{h^{00}} + \frac{\alpha f'(\varphi_0)}{r_{AB}^2} \mathcal{C}_{h^{00}}\right)\right)\\
&+ \mphi \frac{4 G^2 m_A^2 \alpha_A^2}{c^4 r_{AB}}+ \mphi^2 \frac{G^2 m_A^2 }{c^4}(\mathcal{D}_{h^{00}} + \log\frac{r_{AB}}{\mathcal{R}} \mathcal{E}_{h^{00}} + \frac{\alpha f'(\varphi_0)}{r_{AB}}\mathcal{F}_{h^{00}})\\
&+ \mphi^2 \frac{G^2 m_B^2 \alpha_B^2}{c^4} \log\frac{s_B}{\mathcal{R}},
\eal
\ee
with the calligraphic coefficients in \eqref{eq:Coeffh00NZ}.

\begin{align}\label{eq:NZsolhii}
{h^{i(0)}_i}_{\mathcal{N}}(x_B) =& \frac{G^2}{c^4 r_{AB}^2} \mathcal{A}_{h_i^i}  + \frac{4\, G m_A}{c^4 r_{AB}}\, v_{Ai} v_A^i \nonumber  + m_\phi\frac{4\, G^2 \, m_A m_B\, \alpha_A \alpha_B}{c^4 r_{AB}} \nonumber \\[6pt]
& + m_\phi^2\frac{G^2}{c^4} \mathcal{B}_{h_i^i} - m_\phi^2 \frac{G^2 }{c^4} \ln\!\frac{r_{AB}}{R} \mathcal{C}_{h_i^i} - m_\phi^2\frac{G^2 \, m_B^2 \alpha_B^2}{c^4} \ln\!\frac{s_B}{R},
\end{align}
with the calligraphic coefficients in \eqref{eq:CoeffhiiNZ}.

\begin{align}\label{eq:NZsolphi}
\delta\varphi^{(1)}_{\mathcal{N}}(x_B) =& \frac{G^2 m_A m_B}{c^4 r_{AB}^2} \mathcal{A}_{\delta\varphi} + \frac{G m_A \alpha_A}{2\, c^4 r_{AB}} \mathcal{B}_{\delta\varphi} - \frac{G^2 \alpha\, f'(\varphi_0)}{c^4 r_{AB}^4} \mathcal{C}_{\delta\varphi} - \frac{G^2 m_\phi\, m_A}{c^4 r_{AB}} \mathcal{D}_{\delta\varphi} \nonumber \\
& - \frac{G m_A \alpha_A\, m_\phi}{2\, c^4} \mathcal{E}_{\delta\varphi} + \frac{G^2 m_\phi^2\, m_A}{c^4} \mathcal{F}_{\delta\varphi}  + \frac{2\, G^2 m_\phi^2}{c^4} \ln\!\frac{r_{AB}}{R} \mathcal{G}_{\delta\varphi}  - \frac{2\, G^2 m_\phi^2\, m_B^2 \alpha_B}{c^4} \ln\!\frac{s_B}{R} \nonumber\\
& + \frac{G m_A \alpha_A\, m_\phi^2\, r_{AB}}{4\, c^4} \mathcal{H}_{\delta\varphi}+ \frac{2\, G^2 m_\phi^2\, m_A m_B\, \alpha\, f'(\varphi_0)}{c^4 r_{AB}^2}.
\end{align}
with the calligraphic coefficients in \eqref{eq:CoeffphiNZ}.

\subsection{Far zone integration}

\subsubsection{Small scalar mass expansion field integrals}

To find the total field solutions for field point $x$ in the near zone we also need to compute the appropriate far zone contributions to the full integrals \eqref{eq:formalsolhphi}. The source terms in the far zone come from the near zone field solutions evaluated at $x$ in the far zone so we need to obtain those solutions as well. In both cases we cannot use the rewritten scalar Greens function \eqref{eq:FIphiPN} as retardation effects in the far zone are not small, the time dependence cannot be separated from the spatial dependence, which does happen in \eqref{eq:FIphiPN}. We therefore start again at the level of the scalar field equation \eqref{eq:FEphiFS}. The formal solution then becomes

\be
\varphi = -\frac{G}{c^4} \left[\int \frac{\tilde{\mu}_s(ct-|x-x'|,x')}{|x-x'|}d^3x' - \int d^3x' \int_{-\infty}^{t-|x-x'|/c} dt' \frac{\mphi J_1(\mphi \sqrt{(ct-ct')^2-|x-x'|^2})}{\sqrt{(ct-ct')^2-|x-x'|^2}} \tilde{\mu}_s(t',x') \right].
\ee
Here the first integration captures the contributions on the cone and the second integral contributions from inside the past lightcone. Note $h^{\mu\nu}$ only has the former. In the previous section, to obtain the near zone fields evaluated in the near zone, we did not explicitly assume the scalar field mass to be small, but we cut off the expansion in small retardation effects to second order in the scalar field mass to keep the amount of terms tractable. It's favourable to not assume a lengthscale for the scalar field mass as for the radiation one expects non-perturbative effects related to the scale of the Compton wavelength \cite{Alsing:2011er}. However for the next sections we are mainly interested in finding the solutions to the fields to show the cancellation of the near zone boundary dependent terms and to see if there is already a far zone contribution at 1PN. Therefore we argue it suffices for these two sections to work perturbatively in the scalar field mass and make the calculations somewhat more straightforward, but keep in mind these results only hold for small scalar field masses.

\noindent With this assumption we can use the Taylor expansion of the bessel function
\be
J_{\alpha}(x) = \sum_{\ell=0}^{\infty} \frac{(-1)^{\ell}}{\ell! \Gamma(\ell+\alpha+1)} (\frac{x}{2})^{2\ell+\alpha}.
\ee
The scalar field to second order in the scalar mass then becomes
\be\label{eq:FIvarphiFZ}
\varphi = -\frac{G}{c^4}\left[ \int \frac{\tilde{\mu}_s(ct-|x-x'|,x')}{|x-x'|}d^3x' -\frac{1}{2}m_{\varphi}^2\int d^3 x' \int_{-\infty}^{t-|x-x'|/c} dt'\tilde{\mu}_{s}(t',x') +\mathcal{O}(\mphi^3)  \right].
\ee

\subsubsection{FZ field contributions}
In the far zone the source terms only have non-trivial contributions from the pure field dependent terms as the bodies are located in the near zone. With this in mind the scalar field solution simplifies. The scalar field source in the far zone is given by
\be
\tilde{\mu}_s=(1+\frac{1}{2}h)\mu_s + \frac{c^4}{4 \pi G}h^{\mu\nu}\partial_{\mu}\partial_{\nu}\varphi -\frac{c^4}{4 \pi G}\frac{1}{2} m_{\varphi}^2 h \varphi,
\ee
with
\be
\mu_s = -\frac{c^4}{16 \pi G}\alpha f'(\varphi_0) R_{GB}^2.
\ee
Now $h^{\mu\nu}\partial_{\mu}\partial_{\nu}\varphi$ starts at order $\mathcal{O}(c^{-6})$ which is a higher effective PN order than we consider. $R_{GB}$ does have $\mathcal{O}(c^{-4})$ contributions, however all terms are proportional to the product of two second order derivatives to components of $h^{\mu\nu}$, see Appendix A in~\cite{Shiralilou:2021mfl}. The derivatives result in the far zone contributions scaling with higher order in $1/r$ which also raises the effective PN order of these contributions. Therefore the only relevant contribution in the source we have to consider to our order in the expansions is
\be
\tilde{\mu}_s= -\frac{c^4}{4 \pi G}\frac{1}{2} m_{\varphi}^2 h \varphi.
\ee
The second integral of \eqref{eq:FIvarphiFZ} is then higher order in the scalar field mass, hence only the first integration of the far zone scalar field is relevant
\be\label{eq:FZintphifinal}
\varphi = 2\pi m_{\varphi}^2\left[ \int \frac{h \varphi}{|x-x'|}d^3x'  +\mathcal{O}(\mphi^3)  \right],
\ee
which now has the same type of integration as for $h^{\mu\nu}$~\eqref{eq:formalsolhphi}.

\begin{figure}[h!]
\centering
\begin{subfigure}{0.8\textwidth}
   \includegraphics[width=\linewidth]{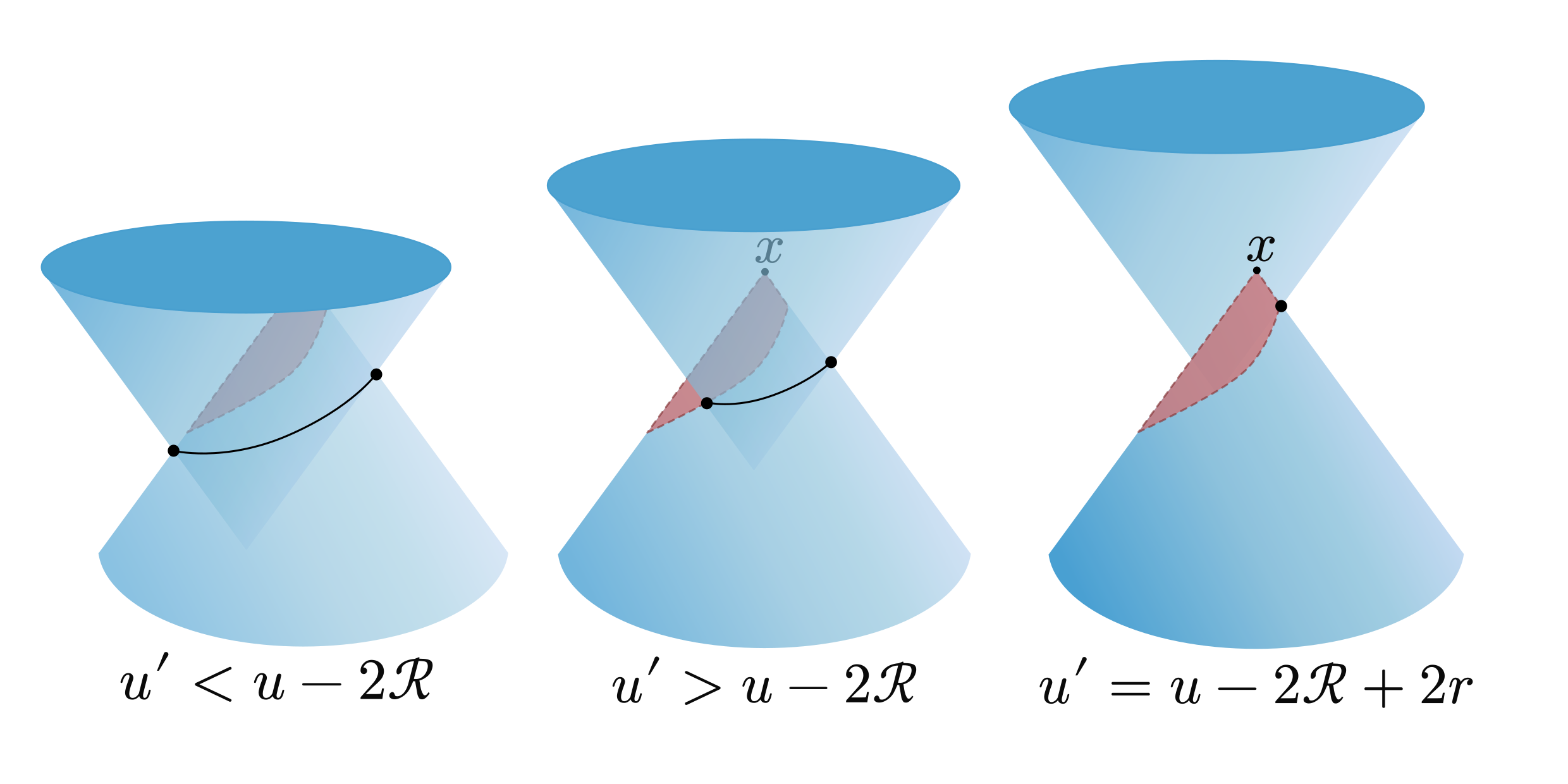}
       \centering
\end{subfigure}
\caption[]{Geometric interpretation of the FZ integration, the future pointed cone intersects the near zone part (red) of the past ligthcone for a certain region of the integration variable $u'$. The angular integration, drawn with the black curve, is then restricted. }
\label{fig:FZboundary}
\end{figure}

\noindent To obtain the far zone portion of the field integrals explicitly, the computation becomes more complicated as the integration domain is over the cone instead of a constant time slice. We follow closely~\cite{Pati:2000vt} to write the integral in a more convenient way. By making the change of variables from $r', \theta', \phi'$ (with $r'=|x'|$) to $u', \theta', \phi'$ by 
\be
u' = ct-r' -|x-x'|,
\ee
the field integral of $h^{\alpha\beta}$ \eqref{eq:formalsolhphi} can be written as
\be
h_{\mathcal{F}}^{\alpha\beta} = \frac{4G}{c^4} \int_{-\infty}^u d u^{\prime} \oint_{S\left(u^{\prime}\right)} \frac{\mu^{\alpha\beta}\left(\left(u^{\prime}+r^{\prime}\right) / c, x^{\prime}\right)}{c t-u^{\prime}-n^{\prime} \cdot x} r^{\prime}\left(u^{\prime}, \Omega'\right)^2 d \Omega^{\prime}.
\ee
With this change of variables the integration can be interpreted as first integrating over the intersection of a future null cone $ct'=u'+r'$ with the past cone followed by the integration over $u'$ such that it's integrated over all such future cones, terminating when the future cone is tangent to the past cone. For a more explicit description of this interpretation and schematic figures of the intersections of the cones, see~\cite{Pati:2000vt}. It's convenient to set the field point $x$ along the z-axis such that $n'\cdot x=r\cos{\theta'}$. One can then write
\be
\bal
h_{\mathcal{F}}^{\alpha\beta} &= \frac{4G}{c^4}\bigg[ \int_{u-2\mathcal{R}}^{u-2\mathcal{R} +2r
}du'\int_0^{2\pi}d\phi'\int_{1-\gamma}^{1} \frac{\mu^{\alpha\beta}\left(\left(u^{\prime}+r^{\prime}\right) / c, x^{\prime}\right)}{c t-u^{\prime}-n^{\prime} \cdot x} r^{\prime}\left(u^{\prime}, \Omega'\right)^2 d\cos{\theta'}\\
&+ \int_{-\infty}^{u-2\mathcal{R} }du'\oint \frac{\mu^{\alpha\beta}\left(\left(u^{\prime}+r^{\prime}\right) / c, x^{\prime}\right)}{c t-u^{\prime}-n^{\prime} \cdot x} r^{\prime}\left(u^{\prime}, \Omega'\right)^2 d^2\Omega' \bigg],
\eal
\ee
where $\gamma = (u-u')(2r-2\mathcal{R} +u -u')/2r\mathcal{R}$. The split in the integration over $u'$ and the incomplete angular integration in the first line correspond to that in the first integration range over $u'$ the future cone intersects the near zone partly which we want to exclude in the far zone calculation. Therefore the angular integration is such that the integration starts from the near zone boundary onward. For the second range of $u'$ there is no near zone crossing and there can be a full angular integration, see Fig.~\ref{fig:FZboundary}.

The expression can be further rewritten by restricting to source terms of the form
\be\label{eq:formsource}
\mu^{\alpha\beta}=\frac{1}{4\pi} f_{B,L}(u') r'^{-B} n'^{<L>},
\ee
with $f$ a function of retarded time $u'=c t'-r'$, exponent $B$ an integer and $n'^{<L>}$ the symmetric trace free (STF) product of $\ell$ unit vectors. The source functions are the near zone field evaluated in the far zone, we derived these expressions up to 1PN in Appendix~\ref{sec:AppC}. The scalar~\eqref{eq:phiNZFZ} and $h^{00}$~\eqref{eq:h00NZFZ} are written in multipole expansions and therefore automatically adopt the form of~\eqref{eq:formsource}. Defining $$z=\mathcal{R}/r,$$ $$\zeta = (ct-u')/r = 1+(u-u')/r,$$ $$y=n\cdot n' = \cos{\theta},$$ implying $$r'=r(\zeta^2-1)/2(\zeta-y).$$ Then we use a change of variables to $\zeta$ and integrate over $\phi'$ using $n'^{<L>} = N_{\ell}\sum_{m=-\ell}^{\ell}\mathcal{Y}_{\ell m}^{<L>}Y_{\ell m}(\theta',\phi')$ with $N_{\ell}=4\pi\ell!/(2\ell+1)!!$ and $\mathcal{Y}_{\ell m}^{<L>}$ a constant STF tensor satisfying $\mathcal{Y}_{\ell,-m}^{<L>}=(-1)^{m}{\mathcal{Y}^{*}_{\ell m}}^{<L>}$. Followed by a Taylor expansion of $f_{B,L}(u') = f_{B,L}(u-r(\zeta-1))$ around $u+r=ct$ for field points in the near zone, we obtain for $B>2$

\be\label{eq:hfarzoneBover2}
\bal
h_{\mathcal{F},B>2}^{\alpha\beta} &= \frac{4 G}{c^4} \frac{1}{2} n^{<L>} \frac{2}{r}^{B-2} \sum_{q=0}^{\infty} r^{q} \frac{d^q f(t)}{dt^q} \frac{(-1)^q}{c^q q!} \bigg[\int_{2z-1}^{1+2z} d\zeta \frac{(\zeta-1)^q}{(\zeta^2 -1)^{B-2}} \int_{1-\gamma}^1 dy P_L(y) (\zeta-y)^{B-3}\\
&+\int_{2z+1}^{\infty} d\zeta \frac{(\zeta-1)^q}{(\zeta^2 -1)^{B-2}} \int_{-1}^1 dy P_L(y) (\zeta-y)^{B-3}\bigg].
\eal
\ee
For the case of $B\leq 2$ the integral of $\zeta$ to infinity becomes divergent, we rewrite the integration then as follows
\be
\bal
h_{\mathcal{F},B<0}^{\alpha\beta} &= \frac{4 G}{c^4} \frac{1}{2} n^{<L>} \frac{2}{r}^{B-2} \sum_{q=0}^{\infty} r^{q} \frac{d^q f(t)}{dt^q} \frac{(-1)^q}{c^q q!}\times\\
&\bigg[\int_{2z-1}^{1+2z} d\zeta \frac{(\zeta-1)^q}{(\zeta^2 -1)^{B-2}} \int_{-1}^1 dy P_L(y) (\zeta-y)^{B-3}\\
&-\int_{2z-1}^{1+2z} d\zeta \frac{(\zeta-1)^q}{(\zeta^2 -1)^{B-2}} \int^{1-\gamma}_{-1} dy P_L(y) (\zeta-y)^{B-3}\\
&+\int_{1}^{\infty} d\zeta \frac{(\zeta-1)^q}{(\zeta^2 -1)^{B-2}} \int_{-1}^1 dy P_L(y) (\zeta-y)^{B-3}\\
&-\int_{1}^{1+2z} d\zeta \frac{(\zeta-1)^q}{(\zeta^2 -1)^{B-2}} \int_{-1}^1 dy P_L(y) (\zeta-y)^{B-3}\bigg].
\eal
\ee
Reversing the boundaries of the last integral and adding it to the first integral we find
\be
\bal
h_{\mathcal{F},B<2}^{\alpha\beta} &= \frac{4 G}{c^4} \frac{1}{2} n^{<L>} \frac{2}{r}^{B-2} \sum_{q=0}^{\infty} r^{q} \frac{d^q f(t)}{dt^q} \frac{(-1)^q}{c^q q!}\times\\
&\bigg[\int_{2z-1}^{1} d\zeta \frac{(\zeta-1)^q}{(\zeta^2 -1)^{B-2}} \int_{-1}^1 dy P_L(y) (\zeta-y)^{B-3}\\
&-\int_{2z-1}^{1+2z} d\zeta \frac{(\zeta-1)^q}{(\zeta^2 -1)^{B-2}} \int^{1-\gamma}_{-1} dy P_L(y) (\zeta-y)^{B-3}\\
&+\int_{1}^{\infty} d\zeta \frac{(\zeta-1)^q}{(\zeta^2 -1)^{B-2}} \int_{-1}^1 dy P_L(y) (\zeta-y)^{B-3}\bigg].
\eal
\ee
The divergent integral we leave unevaluated in the first place and explain below how we treat this piece. The scalar field integral~\eqref{eq:FZintphifinal} can be rewritten in the same way, for which only the factor $4$ in the prefactor of \eqref{eq:hfarzoneBover2} is not there.  \\

\noindent We first look at the far zone contribution to $h^{00(0)}$. Substituting the NZ solutions evaluated in the far zone \eqref{eq:h00NZFZ}, \eqref{eq:phiNZFZ} in the pure field dependent terms in the source \eqref{eq:PNsources}, to leading order in $1/r$ we find the following source term
\be
\mu^{00} = -\frac{G \mphi^2 M_s^2}{16 \pi c^4 r^2} +\mathcal{O}(\frac{1}{r^3}),
\ee
which indeed has the form or \eqref{eq:formsource}. Compared to this form we thus have $f = \frac{-G \mphi^2 M_s^2}{4c^4}$, $B=2$ and $L=0$ and $M_s = \int \tilde{\mu}_s d^{3}x'$ which to leading order in $1/c$ \eqref{eq:PNsources} is given by $M_s=-c^2 m_A \alpha_A-c^2 m_B \alpha_B$. As this is a constant we don't have to Taylor expand $f$ as done in \eqref{eq:hfarzoneBover2}. Then we can write the far zone field as
\be
h_{\mathcal{F}}^{00} = -\frac{G^2 \mphi^2 (m_A \alpha_A +m_B \alpha_B)^2}{c^4} \bigg( -1 + \log(\frac{r}{R}) +\frac{1}{r}\int_0^{\infty} \log(\frac{s+r}{s}) ds \bigg),
\ee
where we did a change of variables of the divergent integral using $s=\frac{r}{2}(\zeta-1)$. This integral looks similar to the gravitational wave tail integrals~\cite{Blanchet:2013haa}. However, in the case of tail contributions to logarithmic term is multiplied with radiative moments of second order or higher. Therefore we cannot use similar argument that the multipole moments stay convergent in the infinite past to ensure the tail integral is finite, see e.g. Sec. 2.4.2 of~\cite{Blanchet:2013haa}. Because of this we extract the infinite constant part after the integration and keep this explicit in the rest of our calculations. We write the integral as 
\be
\bal
\lim\limits_{\epsilon\xrightarrow{}0}\lim\limits_{\Lambda \xrightarrow[]{}\infty} \int_{\epsilon}^{\Lambda} \log(\frac{s+r}{s})ds &=\lim\limits_{\epsilon\xrightarrow{}0}\lim\limits_{\Lambda \xrightarrow[]{}\infty} (r+\Lambda)\log(r+\Lambda)-\Lambda\log(\Lambda)-(r+\epsilon)\log(r+\epsilon) +\epsilon \log(\epsilon)\\
&\sim\lim\limits_{\Lambda \xrightarrow[]{}\infty} r(1-\log(r) +\log(\Lambda)) +\mathcal{O}(\epsilon^2) + \mathcal{O}(\Lambda^2),
\eal
\ee
hence
\be
h_{\mathcal{F}}^{00} = \lim\limits_{\Lambda \xrightarrow[]{}\infty} -\frac{G^2 \mphi^2 (m_A \alpha_A +m_B \alpha_B)^2}{c^4} \bigg(\log(\frac{\Lambda}{R})+\mathcal{O}(\epsilon^2) + \mathcal{O}(\Lambda^2) \bigg) +\mathcal{O}(\frac{1}{r}).
\ee
Comparing this to the $\mathcal{R}$ dependent terms in the near zone solution
\be
\bal
h_{\mathcal{N}}^{00}&\propto  -\frac{G^2 m_\varphi^2}{c^4} \log(\mathcal{R}) \Bigl[
    m_A^2 \alpha_A^2 + 2\, m_A m_B \, \alpha_A \alpha_B
  \Bigr]  - \frac{G^2 m_\varphi^2 \, m_B^2 \alpha_B^2}{c^4} \log(\mathcal{R})\\
  &= -\frac{G^2 \mphi^2 (m_A \alpha_A + m_B \alpha_B)^2}{c^4}\log(\mathcal{R}),
\eal
\ee
we can see that in the total sum the boundary terms vanish. $h_{\mathcal{F}}^{00}$ still contains the divergent term $\lim\limits_{\Lambda \xrightarrow[]{}\infty}\log({\Lambda})$ due to the lightcone extending to infinity. One could argue as physically black holes are formed at a finite time in the past and hence the black hole binaries we want to study go back a large but finite time in the past, then this contribution from the far zone will be finite and constant. Physical observables, for example gravitational radiation, which depends on derivatives to the fields, would not contain this constant contribution. We find the same leading order result for the $h_{\mathcal{F}}^{ii}$ expression. \\

\noindent Similar computation for the FZ scalar field contribution gives
\be
\bal
\varphi_{\mathcal{F}} &= \frac{2 G^2 \mphi^2}{c^4}\left(\alpha_A(m_A^2 + m_A m_B)+\alpha_B(m_B^2+m_A m_B)\right)\\
&\times\bigg( -1 + \log(\frac{r}{R}) +\frac{1}{r}\int_0^{\infty} \log(\frac{s+r}{s}) ds \bigg),
\eal
\ee
again rewriting the diverging integration as above gives 
\be
\varphi_{\mathcal{F}} = \lim\limits_{\Lambda \xrightarrow[]{}\infty} \frac{2 G^2 \mphi^2}{c^4}\left(\alpha_A(m_A^2 + m_A m_B)+\alpha_B(m_B^2+m_A m_B)\right) \bigg(\log(\frac{\Lambda}{\mathcal{R}})+\mathcal{O}(\epsilon^2) + \mathcal{O}(\Lambda^2) \bigg) +\mathcal{O}(\frac{1}{r}).
\ee
Comparing with the near zone boundary dependent terms

\be
\varphi_{\mathcal{N}}\propto \frac{2\, G^2 m_\phi^2}{c^4} \ln\!\mathcal{R} \Bigl[
     m_A^2 \alpha_A
   + m_A m_B \left(\alpha_A + \alpha_B\right)
  \Bigr] + \frac{2\, G^2 m_\phi^2\, m_B^2 \alpha_B}{c^4} \ln\!\mathcal{R},
\ee
we can see the boundary dependent terms cancel. Interestingly we thus already have a constant far zone contributions for the fields at 1PN in this massive extension of the theory which is not present in massless scalar-Gauss-Bonnet. The total field solutions are then given by the near zone plus far zone contributions.

\section{Binary dynamics}\label{sec:BinaryDyn}

Now with the field solutions evaluated at the bodies at hand we can continue analyzing the binary dynamics of the system up to 1PN. Therefore we will substitute our obtained fields in the 1PN expanded total action. \\
The total point particle dynamical two body action is given by the addition of the actions evaluated at point particle A and B respectively. This action in terms of the fields is given by \eqref{eq:SmsGB} plus \eqref{eq:Spp} substituting the expansion of the metric $g_{\mu\nu}$ \eqref{eq:gmunu1PN} and the scalar field \eqref{eq:phi1PN}.

\be\label{eq:Stot}
\begin{aligned}
S = S_{msGB}+S_{pp} =& -c^2\int dt\, m_B \bigg[1-\frac{1}{8} h^{00(0)} + \frac{1}{2}\alpha_B \delta\varphi^{(0)} -\frac{1}{2}\frac{v_B^2}{c^2} -\frac{1}{8} h^{00(1)}-\frac{1}{8} h^{i(0)}_i\\
&+\frac{1}{2}\alpha_B\delta\varphi^{(1)} - \frac{1}{8}\frac{v_B^4}{c^4} +\frac{1}{16}{h^{00(0)}}^2 -\frac{3}{16} h^{00(0)}\frac{v_B^2}{c^2} -\frac{1}{4}\alpha_B\delta\varphi^{(0)}\frac{v_B^2}{c^2} \\
&+ \frac{1}{2}h^{0(0)}_j\frac{v_B^j}{c} +\mathcal{O}(c^{-4})\bigg]_{x\xrightarrow{}x_B} + (B\leftrightarrow A).
\end{aligned}
\ee
From the total action we can obtain the two body Lagrangian via $\mathcal{L}=\frac{dS}{dt}$ from which we can derive the binaries acceleration, center-of-mass (CoM) coordinate transformations and binding energy.

\subsection{Two body lagrangian}
We find the two body lagrangian by substituting the total field solutions in \eqref{eq:Stot}. We can substitute the fields evaluated at $x=x_B$ in \eqref{eq:Stot} to obtain the Lagrangian for body $B$ $\mathcal{L}_B$ and reverse all the labels $(B\leftrightarrow A)$ to obtain $\mathcal{L}_A$. The two body lagrangian is then found by $\mathcal{L} = \mathcal{L}_A + \mathcal{L}_B$. We present the total lagrangian per order in the scalar field mass. 
\be\label{eq:Ltot}
\mathcal{L}_{2body} = \mathcal{L}^{10}+\mphi \mathcal{L}^{11}+\mphi^2\mathcal{L}^{12},
\ee
\begin{align}\label{eq:L10}
\mathcal{L}^{(10)} = \;
& -c^2 (m_A + m_B)  + \frac{m_A}{2} v_{Ai} v_A^i
  + \frac{m_B}{2} v_{Bi} v_B^i
  + \frac{m_A}{8c^2} \bigl(v_{Ai} v_A^i\bigr)^2
  + \frac{m_B}{8c^2} \bigl(v_{Bi} v_B^i\bigr)^2 \nonumber \\[6pt]
& + \frac{G m_A m_B}{r_{AB}} + \frac{G m_A m_B\, \alpha_A \alpha_B}{r_{AB}} + \frac{G m_A m_B}{c^2 r_{AB}}\mathcal{A}_{\mathcal{L}^{(10)}} + \frac{G m_A m_B\, \alpha_A \alpha_B}{c^2 r_{AB}}\mathcal{B}_{\mathcal{L}^{(10)}}\nonumber\\
&- \frac{G^2 m_A m_B (m_A + m_B)}{ c^2 r_{AB}^2}\mathcal{C}_{\mathcal{L}^{(10)}} - \frac{G^2 m_A m_B}{2\, c^2 r_{AB}^2} \mathcal{D}_{\mathcal{L}^{(10)}}+ \frac{G^2 \alpha\, f'(\varphi_0)}{c^2 r_{AB}^4} \mathcal{E}_{\mathcal{L}^{(10)}},
\end{align}
with the calligraphic coefficients in \eqref{eq:CoeffL10}.

\noindent This result coincides with the lagrangian found in \cite{Julie:2019sab,Julie:2019sabErr} for massless sGB theory. For the first and second order scalar mass corrections we find
\begin{align}\label{eq:L11}
\mathcal{L}^{(11)} = \;
& - \frac{1}{2}G m_A^2 \alpha_A^2 +\frac{1}{2}G m_B^2 \alpha_B^2
  -G m_A m_B\, \alpha_A \alpha_B \nonumber  + G\frac{1}{4c^2} \mathcal{A}_{\mathcal{L}^{(11)}} \\
& + G\frac{m_A m_B\, \alpha_A \alpha_B}{2c^2}\mathcal{B}_{\mathcal{L}^{(11)}} + \frac{G^2 m_A m_B}{c^2 r_{AB}} \mathcal{C}_{\mathcal{L}^{(11)}} + \frac{G^2 m_A m_B\, \alpha_A \alpha_B}{c^2 r_{AB}} \mathcal{D}_{\mathcal{L}^{(11)}}\nonumber\\
& + \frac{G^2 m_A m_B(m_A+m_B) \alpha_A^2 \alpha_B^2}{c^2 r_{AB}} + \frac{G^2 m_A m_B}{c^2 r_{AB}} \mathcal{E}_{\mathcal{L}^{(11)}} + \frac{G^2 m_A m_B}{c^2 r_{AB}}\mathcal{F}_{\mathcal{L}^{(11)}},
\end{align}
with the calligraphic coefficients in \eqref{eq:CoeffL11}.

\begin{align}\label{eq:L12}
\mathcal{L}^{(12)} = \;
& \frac{G m_A m_B\, \alpha_A \alpha_B\, r_{AB}}{2} + \frac{G m_A m_B\, \alpha_A \alpha_B\, r_{AB}}{4 c^2} \mathcal{A}_{\mathcal{L}^{(21)}}+\frac{G^2 m_A m_B}{c^2}\mathcal{B}_{\mathcal{L}^{(21)}}\\
& - \frac{G^2 m_A m_B\, \alpha_A \alpha_B}{c^2} \mathcal{C}_{\mathcal{L}^{(21)}}  + \frac{G^2}{c^2} \ln\!\frac{r_{AB}}{\Lambda} \mathcal{D}_{\mathcal{L}^{(21)}}  - \frac{G^2}{c^2} \Bigl[
    m_A^3 \alpha_A^2 \ln\!\Lambda
    + m_B^3 \alpha_B^2 \ln\!\Lambda
  \Bigr] \nonumber \\
& + \frac{G^2 \alpha\, f'(\varphi_0)}{c^2 r_{AB}^2}\mathcal{E}_{\mathcal{L}^{(21)}},
\end{align}

with the calligraphic coefficients in \eqref{eq:CoeffL12}.

\noindent We compare these results to the two body lagrangian obtained for massive scalar tensor theory with an EFT approach in \cite{Diedrichs:2023foj}. This work obtains the expressions unperturbed in the scalar field mass. The lagrangian expressions in eq (62)-(64) at first and second order expanded in the scalar field mass can be compared to the expressions above for $\alpha=0$. We should take into account that they didn't work perturbatively in the mass which can lead to capturing non-perturbative effects or consequences for the regularization which we do not capture. Furthermore they note they do not include self-force diagrams, these contributions correspond to the linear in scalar mass terms that survive the PF operation, e.g. the last term of the scalar field 0PN solution \eqref{eq:sol0PNh00phi}. They note these constant terms can be absorbed by redefining the masses and scalar charges. We therefore give the conversion of their mass and scalar charge related parameters to our convention and the shift for the mass and scalar charge terms for these self-force contributions become explicit and compare their expressions to ours after these transformations.

\begin{align}
&q_2 = \frac{m_B \alpha_B}{2\sqrt{2}}, \hspace{2.15cm} q_1 = \frac{m_A \alpha_A}{2\sqrt{2}},\\
&p_2 = \frac{q_2^2}{2 m_B } + \frac{m_B \beta_B}{16 },\quad p_1 = \frac{q_1^2}{2 m_A} + \frac{m_A \beta_A}{16 },\\
&M_1=m_A, \hspace{2.5cm} M_2=m_B .
\end{align}
After these transformations we can make the self-force contributions explicit with the following transformations

\begin{align}
m_A &\xrightarrow{} m_A+\mphi\frac{ G m_A^2 \alpha_A^2 }{ c^2} + \mphi^2 \frac{G^2 m_A^3 \alpha_A^2}{c^4}\left(\frac{\alpha_A^2}{2} + \frac{\beta_A}{2}\right),\\
m_B &\xrightarrow{} m_B+ \mphi\frac{ G m_B^2\alpha_B^2}{ c^2}+ \mphi^2 \frac{G^2 m_B^3 \alpha_B^2}{c^4}\left(\frac{\alpha_B^2}{2} + \frac{\beta_B}{2}\right),\\
\alpha_A &\xrightarrow{}\alpha_A+\mphi \left(\frac{\alpha_A^3 G m_A}{2 c^2}+\frac{\alpha_A
   \beta_A G m_A}{c^2}-\frac{\gamma_E  \alpha_A G m_A}{c^2}\right),  \\
   \alpha_B &\xrightarrow{}\alpha_B+\mphi \left(\frac{\alpha_B^3 G m_B}{2 c^2}+\frac{\alpha_B
   \beta_B G m_B}{c^2}-\frac{\gamma_E  \alpha_B G m_B}{c^2}\right).
\end{align}

\noindent Substituting this in eq (62)-(64) of \cite{Diedrichs:2023foj}, expanding to second order in the scalar field mass, we first compare the terms linear in the scalar field mass to \eqref{eq:L11} and indeed find their difference to vanish. 
\be
\mathcal{L}^{(11)}_{\textrm{\cite{Diedrichs:2023foj}}}-\mathcal{L}^{(11)}=0.
\ee
At second order in the scalar field mass we find the following difference with \eqref{eq:L12}

\begin{align}\label{eq:diffL12}
\mathcal{L}^{(12)}_{\textrm{\cite{Diedrichs:2023foj}}}-\mathcal{L}^{(12)} &= 2 \alpha_A\alpha_B \frac{G^2 m_A m_B}{c^2}(m_A + m_B )(\gamma_E +\log(2 \mphi) + \log(\Lambda))\nonumber\\
&+\frac{G^2 m_A m_B}{c^2}(m_A \alpha_A^2 + m_B\alpha_B^2 )(\gamma_E +\log(2 \mphi) + \log(\Lambda) )\nonumber\\
&+\frac{G^2 m_A m_B}{c^2}\left(\frac{m_A^2 \alpha_A^2}{m_B}\log(\Lambda)+\frac{m_B^2 \alpha_B^2}{m_A}\log(\Lambda)\right).
\end{align}

\noindent These terms relate to the difference in renormalization schemes. Our obtained Lagrangian in harmonic coordinates at 1PN order is still conservative, no linear dissipative terms are present. The center of mass transformation rules and equations of motion in the subsequent section will be derived from this two body Lagrangian above, depending on the positions and velocities of the two bodies. 

\subsection{Center of mass transformation rules}
The center of mass position vector $\mathbf{G}$ can be derived from the noetherian symmetry of the lagrangian under boost transformations~\cite{deAndrade:2000gf}. The variation to an infinitesimal boost of the lagrangian should take the form of a total derivative so the dynamics remains unchanged. To linear order in the boost velocity $V^i$ the variation to the particles position is given by
\be\label{eq:varposboost}
\delta x_A^i = -V^i t - \frac{1}{c^2}V^j x_A^j v_A^i + \mathcal{O}(V^iV_i).
\ee
The variation to the lagrangian should have the form 
\be
\delta\mathcal{L} = V^i \frac{d Z^i}{dt} + + \mathcal{O}(V^iV_i),
\ee
with $Z^i$ a functional depending on the positions and velocities. The center-of-mass position vector is then given by~\cite{deAndrade:2000gf}
\be\label{eq:defG}
G^i = -Z^i + \bigg(\frac{1}{c^2} x_A^i p_A^j v_A^j + \AB\bigg),
\ee
with $p_A^i = \frac{\delta \mathcal{L}}{\delta v_A^i} $
this vector is conserved in the frame where the canonical momentum vanishes. \\

\noindent We varied the lagrangian \eqref{eq:Ltot} with respect to \eqref{eq:varposboost}. To write the variation to the lagrangian as a total time derivative we specialize to circular orbits and parametrize the orbit as 
\begin{align}
\mathbf{x_A} = \begin{bmatrix}
    |x_A|\cos(\omega t) \\
    |x_A| \sin(\omega t) \\
    0
\end{bmatrix}, \quad \mathbf{x_B} = \begin{bmatrix}
    |x_B|\cos(\omega t+\pi) \\
    |x_B| \sin(\omega t+\pi) \\
    0
\end{bmatrix},
\end{align}
and integrated the variation of the lagrangian to time to obtain
\be
Z^i = \frac{1}{V^i} \int \delta\mathcal{L} dt,
\ee
given by
\begin{align}
    Z^i &= -m_A x_A^i +\frac{G m_A m_B}{2 r_{AB} c^2 }\bigg((1+\alpha_A\alpha_B)x_A^i + \frac{|x_A|^2\omega^2 x_A^i}{m_B} - \mphi\frac{m_A\alpha_A^2}{m_B}x_A^i r_{AB} - \mphi\alpha_A \alpha_B x_A^i r_{AB}\nonumber\\
    &+ \frac{\mphi^2}{2} \alpha_A\alpha_B (|x_A|+|x_B|)^2 x_A^i \bigg) + \AB + \mathcal{O}(c^{-4}).\nonumber
\end{align}
Formally there is an integration constant as well reflecting the freedom in choice of origin.
After substituting in \eqref{eq:defG} and varying the lagrangian to the velocity to obtain $p^i$ the center of mass position vector is given by
\begin{align}\label{eq:COMposG}
G^i &= x_A^i \bigg[m_A  -\frac{G m_A m_B}{2 r_{AB} c^2 }\bigg((1+\alpha_A\alpha_B) - \frac{|x_A|^2\omega^2 }{2 m_B} - \mphi\frac{m_A\alpha_A^2}{m_B} r_{AB} - \mphi\alpha_A \alpha_B  r_{AB}\\
    &+ \frac{\mphi^2}{2} \alpha_A\alpha_B (|x_A|+|x_B|)^2  \bigg)\bigg] + \AB + \mathcal{O}(c^{-4}).\nonumber
\end{align}
In the center of mass frame this vector vanishes, in relative coordinates we can write 
\begin{align}\label{eq:COMxA}
x_A^i &=r_{AB}^i \bigg[\frac{m_B}{M} + \frac{m_A m_B}{2 M^3 c^2}(m_A -m_B) v_{AB}^j{v_{AB}}_j \\
&\frac{G m_A m_B}{2 M^3 c^2 r_{AB}}(-m_A^2(1+\alpha_A \alpha_B) + m_B^2(1+\alpha_A \alpha_B) ) +\nonumber\\
&\mphi \frac{G m_A m_B}{2 M^3 c^2 r_{AB}} (-m_A m_B (\alpha_A^2-\alpha_B) + m_A^2\alpha_A\alpha_B - m_B^2 \alpha_A \alpha_B +m_B^2 \alpha_B^2)+\nonumber\\
&\mphi^2\frac{G m_A m_B\alpha_A \alpha_B r_{AB}}{4 M^3 c^2}(-m_A^2 +m_B^2)\bigg],\nonumber
\end{align}
\begin{align}\label{eq:COMxB}
x_B^i &=r_{AB}^i \bigg[-\frac{m_B}{M} + \frac{m_A m_B}{2 M^3 c^2}(m_A -m_B) v_{AB}^j{v_{AB}}_j \\
&\frac{G m_A m_B}{2 M^3 c^2 r_{AB}}(-m_A^2(1+\alpha_A \alpha_B) + m_B^2(1+\alpha_A \alpha_B) ) +\nonumber\\
&\mphi \frac{G m_A m_B}{2 M^3 c^2 r_{AB}} (-m_A m_B (\alpha_A^2-\alpha_B) + m_A^2\alpha_A\alpha_B - m_B^2 \alpha_A \alpha_B +m_B^2 \alpha_B^2)+\nonumber\\
&\mphi^2\frac{G m_A m_B\alpha_A \alpha_B r_{AB}}{4 M^3 c^2}(-m_A^2 +m_B^2)\bigg],\nonumber
\end{align}
where we used $M=m_A+m_B$ and $r_{AB}^2\omega^2 = v_{AB}^j{v_{AB}}_j$.\\

\noindent With the same transformation rules we introduced in the previous section to compare our results to the results of \cite{Diedrichs:2023foj} we compared these transformation rules to their equation (71) and find the expressions coincide as we didn't obtain an $\alpha$ dependent term. This is different from the transformation expressions found in \cite{Shiralilou:2021mfl} in the massless scalar field limit which did obtain an $\alpha$ dependent contribution. However only with setting this contribution to zero do these coordinate transformations let the canonical momentum vanish as the lagrangian does not have velocity dependent $\alpha$ terms they are not present in the momentum expressions.

\subsection{Equation of motion}
From the Euler-Lagrange equations with the two body lagrangian \eqref{eq:Ltot} we obtain the relative equations of motion. Using the center of mass transformations \eqref{eq:COMxA},\eqref{eq:COMxB} we present here below the relative acceleration per order in the scalar field mass.

\begin{align}\label{eq:a10}
{a^{(10)}_{rel}}_k = \;
& - \frac{G(m_A + m_B)}{r_{AB}^2}\bigl(1 + \alpha_A \alpha_B\bigr)\, n_{AB\,k}  + \frac{G^2 n_{AB\,k}}{c^2 r_{AB}^3} \mathcal{A}_{a_{rel}^{(10)}} + \frac{G^2 \alpha_A \alpha_B\, n_{AB\,k}}{c^2 r_{AB}^3}\mathcal{B}_{a_{rel}^{(10)}}\nonumber \\
& + \frac{2\, G^2 m_A m_B\, \alpha_A^2 \alpha_B^2}{c^2 r_{AB}^3}\, n_{AB\,k} + \frac{G^2 (m_A + m_B)\, n_{AB\,k}}{c^2 r_{AB}^3}\mathcal{C}_{a_{rel}^{(10)}}  - \frac{G^2 \alpha\, f'(\varphi_0)\, n_{AB\,k}}{c^2 r_{AB}^5}\mathcal{D}_{a_{rel}^{(10)}} \nonumber \\[6pt]
& - \frac{G\mathcal{E}_{a_{rel}^{(10)}}}{c^2 M^2 r_{AB}^2}
    \, n_{AB\,k}\, v_{AB}^i v_{ABi} + \frac{G\bigl(m_A^3 + m_B^3\bigr)\alpha_A \alpha_B}{c^2 M^2 r_{AB}^2}
    \, n_{AB\,k}\, v_{AB}^i v_{ABi}\nonumber \\
    &+ \frac{3\, G m_A m_B (m_A + m_B)}{2\, c^2 M^2 r_{AB}^2}
    \bigl(1 + \alpha_A \alpha_B\bigr)
    \, n_{AB}^i n_{AB}^j n_{AB\,k}\, v_{ABi} v_{ABj} \nonumber \\[6pt]
& + \frac{4\, G\bigl(m_A^3 + m_B^3\bigr) + 10\, G m_A m_B(m_A + m_B)}{c^2 M^2 r_{AB}^2}
    \, n_{AB}^i v_{ABi}\, v_{AB\,k} \nonumber \\
& - \frac{2\, G m_A m_B (m_A + m_B)\, \alpha_A \alpha_B}{c^2 M^2 r_{AB}^2}
    \, n_{AB}^i v_{ABi}\, v_{AB\,k}, 
\end{align}
with the calligraphic coefficients in \eqref{eq:Coeffarel10}.
\begin{align}\label{eq:a11}
{a^{(11)}_{rel}}_k = \;
& - \frac{G^2 n_{AB\,k}}{c^2 r_{AB}^2}\mathcal{A}_{a_{rel}^{(11)}} - \frac{G^2 m_A m_B\, n_{AB\,k}}{c^2 r_{AB}^2}\, \alpha_A \alpha_B - \frac{G^2 n_{AB\,k}}{c^2 r_{AB}^2} \mathcal{B}_{a_{rel}^{(11)}}\nonumber \\
& - \frac{G^2 m_A m_B\, n_{AB\,k}}{c^2 r_{AB}^2}\, \alpha_A^2 \alpha_B^2 - \frac{G^2 \varsigma\, n_{AB\,k}}{c^2 r_{AB}^2}\mathcal{C}_{a_{rel}^{(11)}} - \frac{G^2 n_{AB\,k}}{c^2 r_{AB}^2} \mathcal{D}_{a_{rel}^{(11)}},
\end{align}
with the calligraphic coefficients in \eqref{eq:Coeffarel11}.
\begin{align}\label{eq:a12}
{a^{(12)}_{rel}}_k = \;
& \frac{1}{2}\,G(m_A + m_B)\,\alpha_A \alpha_B\, n_{AB\,k}  + \frac{G^2 n_{AB\,k}}{c^2 r_{AB}} \mathcal{A}_{a_{rel}^{(12)}}  + \frac{G^2 \alpha\, f'(\varphi_0)\, n_{AB\,k}}{c^2 r_{AB}^3} \mathcal{B}_{a_{rel}^{(12)}} \nonumber \\[6pt]
& - \frac{G(m_A^3 + m_B^3)\,\alpha_A \alpha_B}{2\,c^2 M^2}
    \, n_{AB\,k}\, v_{AB}^i v_{ABi}  - \frac{G m_A m_B(m_A + m_B)\,\alpha_A \alpha_B}{4\,c^2 M^2}
    \, n_{AB}^i n_{AB}^j n_{AB\,k}\, v_{ABi} v_{ABj} \nonumber \\[6pt]
& + \frac{G m_A m_B(m_A + m_B)\,\alpha_A \alpha_B}{c^2 M^2}
    \, n_{AB}^i v_{ABi}\, v_{AB\,k},
\end{align}
with the calligraphic coefficients in \eqref{eq:Coeffarel12}.
\subsection{Binding energy}
The binding energy of the system is related to the two body lagrangian via a legendre transformation
\be\label{eq:LegendreTrEb}
E_b = \frac{\p \mathcal{L}}{\p v_A^i}v_A^i+ \frac{\p \mathcal{L}}{\p v_B^i}v_B^i - \mathcal{L}.
\ee
We would like to express the binding energy in terms of the orbital frequency instead of the coordinate dependent orbital radius. We'll first do this for circularized orbits and subsequently generalize to eccentric orbits.

\subsubsection{Binding energy for circular orbits}
When assuming circular orbits we can write $\dot{r}=\ddot{r}=0$ and for the orbital frequency $\omega^2=-a_i \frac{r_{AB}^i}{r_{AB}^2}$ and thus for the orbital velocity $v^2 = -a_i r_{AB}^i$. We invert this expression per order in $c$ and $\mphi$ to find $r(x)$ introducing the PN parameter 
\be
x=\left(\frac{G (1+\alpha_A \alpha_B) M \omega}{c^3}\right),
\ee
we obtained per order in $\mphi$

\begin{align}\label{eq:r10}
    r(x)^{(10)} &= \frac{G }{(1+\alpha_A \alpha_B) c^2 M} \left(\mathcal{A}_{r^{(10)}}+ \frac{1}{x}\mathcal{B}_{r^{(10)}} \right)+ x^2 \frac{4 c^2}{3 G (1+\alpha_A\alpha_B)^3 G M^2}\alpha f'(\varphi_0)\mathcal{C}_{r^{(10)}},
\end{align}

with the calligraphic coefficients in \eqref{eq:Coeffr10}.

\begin{align}\label{eq:r11}
r(x)^{(11)} &= \frac{G^2 }{3 c^4 x}\mathcal{A}_{r^{(11)}},
\end{align}

with the calligraphic coefficient in \eqref{eq:Coeffr11}.

\begin{align}\label{eq:r12}
r(x)^{(12)} &= \frac{G^3}{c^6} \left[ \frac{1}{x^3}\mathcal{A}_{r^{(12)}} +\frac{1}{x^2}\mathcal{B}_{r^{(12)}} \right] -\frac{G}{c^2 (1+\alpha_A \alpha_B)^2}\alpha f'(\varphi_0) \mathcal{C}_{r^{(12)}},
\end{align}
with the calligraphic coefficients in \eqref{eq:Coeffr12}.

\noindent Inserting the circular orbit conditions and the above expression for $r(x)$ in the binding energy as function of $r_{AB}$, obtained via \eqref{eq:LegendreTrEb}, and truncating the expression at 1PN and second order in the scalar field mass gives

\begin{align}\label{eq:E10}
E^{(10)} = \;
& - \frac{1}{2} \frac{c^2 m_A m_B}{M}x -\frac{c^2 m_A m_B}{24\,M^3\,(1+\alpha_A\alpha_B)^2} x\,\mathcal{A}_{E^{(10)}} + \frac{5\,c^6 m_A m_B\, x^4\,\alpha\, f'(\varphi_0)}{
    3\,G^2\,M^4\,(1+\alpha_A\alpha_B)^4}\mathcal{B}_{E^{(10)}},
\end{align}
with the calligraphic coefficients in \eqref{eq:CoeffE10}.

Which agrees with the result obtained for massless Gauss-Bonnet theory \cite{Shiralilou:2021mfl}.

\begin{align}\label{eq:E11}
E^{(11)} &=\frac{1}{12 (\alpha_A \alpha_B+1) (m_A+m_B)^2}G  x \mathcal{A}_{E^{(11)}}+\nonumber\\
   &\frac{1}{12 (\alpha_A \alpha_B+1) (m_A+m_B)^2}G \mathcal{B}_{E^{(11)}},
\end{align}
with the calligraphic coefficients in \eqref{eq:CoeffE11}.

\be\label{eq:E12}
\bal
E^{(12)} &=\frac{1}{288 (\alpha_A \alpha_B+1)^3
   (m_A+m_B)^2}\alpha  c^2 m_A m_B x^2 f'(\varphi_0) \mathcal{A}_{E^{(12)}}\\
   &+\frac{1}{288 c^2 x (\alpha_A \alpha_B+1)^3 (m_A+m_B)^2}\mathcal{B}_{E^{(12)}}+\frac{1}{288 c^2 (\alpha_A \alpha_B+1)^3 (m_A+m_B)^2}\mathcal{C}_{E^{(12)}}, 
\eal
\ee
with the calligraphic coefficients in \eqref{eq:CoeffE12}.

\noindent We find here a different result for the expression at first and second order in the scalar field mass compared to \cite{Diedrichs:2023foj}. At this stage of the computation on top of the PN expansion in \cite{Diedrichs:2023foj} the energy is expanded to first order in a combinatorial parameter depending on the scalar charges and masses. In the case above we only expanded in PN and to second order in the scalar field mass. \\

\noindent To get more insight on the scales of the different scalar mass dependent terms in the binding energy, we computed the different components for fiducial binary systems as function of the orbital frequency. We looked at a black hole binary systems of total mass $M=m_A+m_B=100M_{\odot}$ and mass ratios $q=\frac{m_A}{m_B}$ of respectively $q=1,\frac{1}{10},\frac{1}{100}$. We specify to a dilatonic coupling function $f(\varphi)=\frac{1}{4}e^{2\varphi}$ and the coupling around the current observational constraint value $\sqrt{\alpha}=1 \textrm{km}$. Furthermore we set as a small scalar mass value $\tilde{m}=0.003$. Formally the mass and $\alpha_{A/B}$ and $\beta_{A/B}$ coefficients need to be matched to the full black hole solution in massive sGB in the asymptotic limit. However it turns out that the matching for the scalar field, which determines the $\alpha_{A/B}$ and $\beta_{A/B}$ is more complex as for the massless scalar field~\cite{Julie:2019sab}. The coefficients can be linked to the coefficient multiplying the exponential fall off of the scalar. However this quantity is not coordinate independent and ambiguities arise which is not the case for the massless scalar. We leave this discussion to the work in preperation~\cite{Diedrichs:2026inprep}. As at this stage we're mostly interested in qualitative features we set the coefficients to their massless values~\cite{Julie:2019sab} $\alpha_{A/B}=-\alpha f'(\varphi_0)/2 m_{A/B}^2$, $\beta_{A/B}=-\alpha f''(\varphi_0)/2 m_{A/B}^2$. \\

\begin{figure}[h!]
\centering
\begin{subfigure}{0.8\textwidth}
   \includegraphics[width=\linewidth]{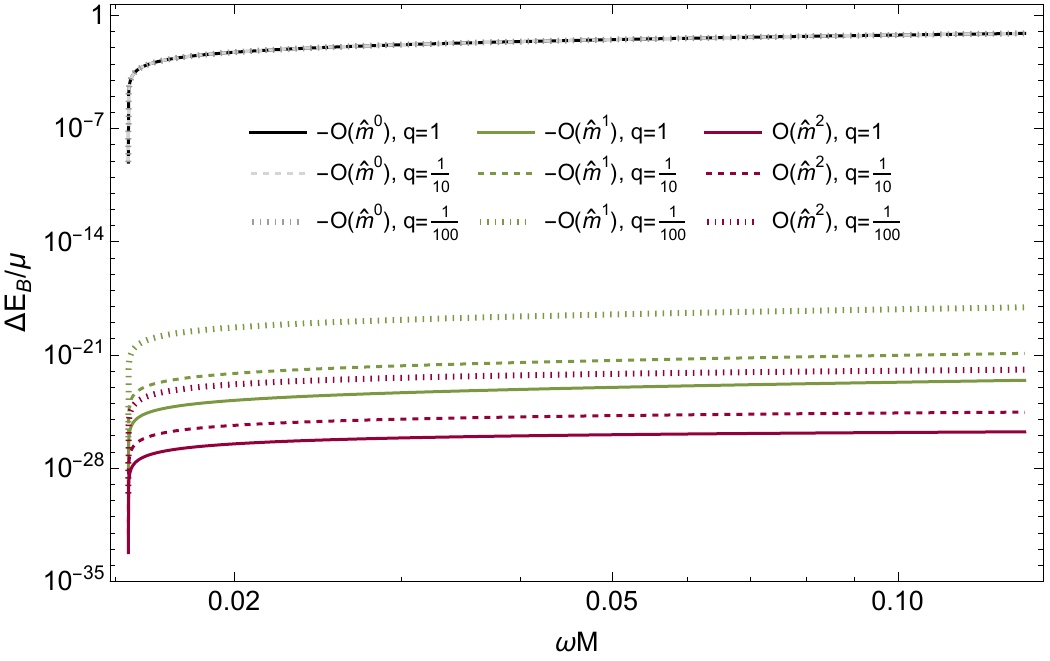}
       \centering
\end{subfigure}
\caption[]{Accumulated binding energy at zeroth (black and grey), first (green) and second (pink) order in the scalar field mass against the orbital frequency. For a system of $M=100 M_{\odot}$ the different mass ratios $q=1,\frac{1}{10},\frac{1}{100}$ are denoted with full, dashed and dotted curves respectively. Note the black/grey and green curves are multiplied with $-1$.}
\label{fig:Ebcircl}
\end{figure}
\noindent In Fig.~\ref{fig:Ebcircl} we show the accumulated binding energy per order in the scalar field mass as a dimensionless quantity divided by $\mu=m_A m_B/(m_A+m_B)$ against the dimensionless orbital frequency. We can see that the scalar mass contributions are many orders of magnitudes less than the massless contributions. A more uneven mass system increases the magnitude of the binding energy for the scalar mass contributions whereas for the point particle+massless contribution this effects is much smaller. From $q=1$ to $q=\frac{1}{10}$ the linear in scalar mass contribution increases order $\mathcal{O}(10^{2})$ and the quadratic order contribution increases $\mathcal{O}(10^{1})$.\\
The curves are plotted as the accumulated binding energy compared to the value at the starting frequency. In this way the unknown $\log(\Lambda)$ dependence in the quadratic scalar mass contribution cancels. However the sign of the curves now only show if the contribution is in- or decreasing with the frequency. The sign of the bare contribution is interesting to know to say if the system becomes more or less bound. We gave the sign of the bare contributions in Table~\ref{tab:Ebsigns}. We can see that linear in mass contribution makes the system less bound while the quadratic contribution increases the bound due to the negative sign. However the sign of the quadratic contribution can change depending on the value for $\log(\Lambda)$.\\
From Fig.~\ref{fig:Ebcircl} we cannot say much about the expectations on how the mass terms contribute to observational quantities in the radiation. The actual frequency evolution is only obtained from equating the reduction in binding energy to the gravitational and scalar energy losses. \\
\begin{table}[h!]
\centering
\begin{tabular}{lllll}
\cline{1-4}
\multicolumn{1}{|l|}{} &
  \multicolumn{1}{l|}{$m_{\varphi}^0$} &
  \multicolumn{1}{l|}{$m_{\varphi}^1$} &
  \multicolumn{1}{l|}{$m_{\varphi}^2$} &
   \\ \cline{1-4}
\multicolumn{1}{|l|}{sign $E_B$ contributions} &
  \multicolumn{1}{l|}{$-$} &
  \multicolumn{1}{l|}{$+$} &
  \multicolumn{1}{l|}{$-$ ($+ \log(\Lambda)$ contribution)} &
   \\ \cline{1-4}
\end{tabular}
\caption[]{Signs of the different contribution to the binding energy.}\label{tab:Ebsigns}
\end{table}

\noindent With this analysis we can quantify the range of validity of our perturbative expansion in the scalar field mass for this frequency range, namely for which values of the scalar field mass per choice of the binary system the quadratic contribution is smaller than the linear contribution.\\ 

\begin{figure}[h!]
\centering
\begin{subfigure}{0.6\textwidth}
   \includegraphics[width=\linewidth]{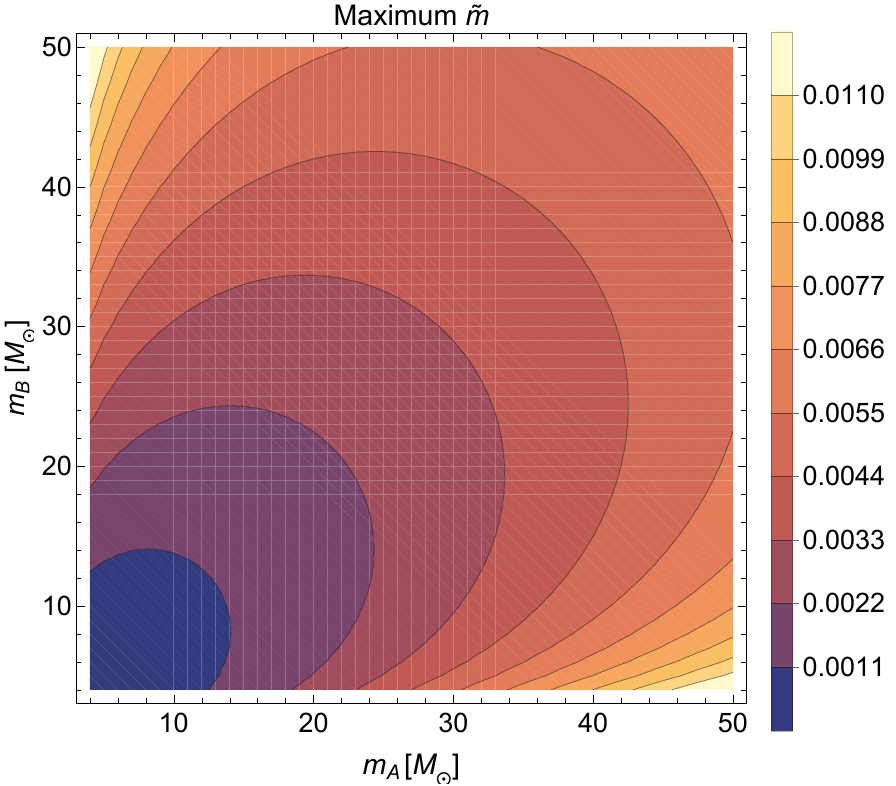}
       \centering
\end{subfigure}
\caption[]{Contourplot showing the maximum value for $\tilde{m}$ for $\sqrt{\alpha}=1\textrm{km}$ the systems for different black hole masses are allowed to have such that the quadratic scalar field contributions in the binding energy stay smaller than the linear contributions.}
\label{fig:maxmhat}
\end{figure}
\noindent The contours in Fig.~\ref{fig:maxmhat} show this maximum value per choice of black hole masses. We see a symmetric dependence on the individual masses as the dimensionless scalar mass is rescaled with the total mass.

\subsubsection{Binding energy eccentric orbits}
Instead of assuming quasi-circular orbits where the timescale for which the separation of the bodies decreases due to energy loss of the radiation is much larger than one orbital period, we can extend to generic orbits by reparametrizing the orbit allowing for eccentricity. The loss in binding energy and GW fluxes back reacting on the orbital parameters are then only defined in an averaged sense over a cycle of radial motion. The derivation below is based on the approach in \cite{Hinderer:2017jcs}. As the expressions are very long we present here the equations as generic functions of the variables and give the final result for the binding energy in terms of the orbital parameters in a separate Mathematica notebook.\\
The generic orbits in the orbital plane can be parametrized as  
\be
\mathbf{x}=r(\cos(\phi),\sin(\phi),0),
\ee
with $r$ the orbital separation and $\phi$ the phase angle. This gives for the norm of the velocity
\be
v^2=\dot{r}^2+r^2\dot{\phi}.
\ee

\noindent The binding energy we derived in \eqref{eq:LegendreTrEb} before specifying to circular orbits depends on $r$ and $v$, we replace the velocities with the parametrization above resulting in
\be\label{eq:Ebparam} 
E_b = f(\dot{r}[t],r[t],\dot{\phi}[t])
\ee

\noindent We can also find an expression for the angular momentum of the system along the $z$-axis $L_z$ by differentiating the two-body lagrangian~\eqref{eq:Ltot} in CM coordinates~\eqref{eq:COMxA},\eqref{eq:COMxB} and parametrized velocity to $\dot{\phi}[t]$ finding the canonical momentum related to the angular variable. We rewrite this equation to find expression for $\dot{\phi}[t]$ in terms of $L_z$
\be\label{eq:phiprimeLz}
\dot{\phi}[t] \propto L_z + g(r[t],\dot{r}[t]).
\ee

\noindent We now find the turning points of the motion defined by $\dot{r}=0$. We substitute \eqref{eq:phiprimeLz} in \eqref{eq:Ebparam}, expand the equation to first order in $1/c^2$ and second order in the scalar field mass and solve for $r[t]$ per order in the expansion, this results in the solution of two turning points 
\be\label{eq:rturningEbLz}
\bal
r_1=&h(E_b, L_z),\\
r_2=&k(E_b,L_z).
\eal
\ee
For eccentric orbits we define the orbital parameters $e$, $p$ and parametrize the turning points as 
\be\label{eq:rtunringpe}
\bal
r_1=&\frac{p}{1-e},\\
r_2=&\frac{p}{1+e}.
\eal
\ee
We then set our obtained \eqref{eq:rturningEbLz} to the parametrization \eqref{eq:rtunringpe} to solve per order in $c$ and $\mphi$ for the binding energy and angular momentum in terms of $e$ and $p$
\be\label{eq:EbLzinep}
\bal
E_b=&l(e,p),\\
L_z=&m(e,p).
\eal
\ee
The full radial function can be parametrized by introducing an angular variable $\xi$
\be\label{eq:seprparam}
r = \frac{p}{1+e \cos(\xi)},
\ee
which has the form of an ellipse equation with $p$ the role of the semi-latus rectum, $e$ the eccentricity and $\xi$ the true anomaly. The turning points correspond to $\xi=(0,\pi)\mathrm{mod}(2\pi)$. With this parametrization we can define the radial and azimuthal velocities in terms of the orbital elements. First we substitute \eqref{eq:EbLzinep}, \eqref{eq:phiprimeLz} and \eqref{eq:seprparam}in \eqref{eq:Ebparam} and solve per order in $c$ and $\mphi$ for $\dot{r}^2$. We obtain $r^2\dot{\phi}^2$ in terms of the orbital variables by substituting \eqref{eq:seprparam} and \eqref{eq:phiprimeLz}. Thus we have 
\be\label{eq:rdotphidotinpe}
\bal
\dot{r}^2 =& n(p,e,\xi),\\
r^2\dot{\phi}^2=&q(p,e,\xi),
\eal
\ee
and hence 
\be
v^2 = \dot{r}^2 + r^2\dot{\phi}^2 = s(p,e,\xi).
\ee
We can substitute these expressions in the obtained binding energy to find it in the orbital parameters $e, p, \xi$. The full expression for $E_b(e, p, \xi)$ can be found in the \texttt{Mathematica} notebook via \texttt{Github}~\cite{githublink}.\\

\noindent Differentiating \eqref{eq:seprparam} gives the equation of motion for $\xi$
\be
\dot{\xi}=\frac{(1+e \cos \xi)^2}{e p \sin \xi} \dot{r}+\frac{\cot \xi}{e} \dot{e}-\frac{1+e \cos \xi}{e p \sin \xi} \dot{p}.
\ee
The equations of motions for $e$ and $p$ can be obtained from $E_b$ and $L_z$ \eqref{eq:EbLzinep} with the transformation
\be
\dot{e}=c_{E p} \frac{d L_z}{d t}-c_{L p} \frac{d E}{d t}, \quad \dot{p}=c_{L e} \frac{d E}{d t}-c_{E e} \frac{d L_z}{d t},
\ee
with coefficients
\be
c_{C b}=\frac{\partial C / \partial b}{(\partial E / \partial p)\left(\partial L_z / \partial e\right)-(\partial E / \partial e)\left(\partial L_z / \partial p\right)} .
\ee

\noindent Renaming the conservative dynamics of the radial phase $\mathcal{P}$ 
\be
\dot{\xi}=\mathcal{P}(p,e,\xi)+\frac{\cot \xi}{e} \dot{e}-\frac{1+e \cos \xi}{e p \sin \xi} \dot{p},
\ee
which we can obtain by equating the time derivative of \eqref{eq:seprparam} for constant (e,p) to \eqref{eq:rdotphidotinpe} and solving for $\dot{\xi}$. One can only solve for the full set of orbital parameters when having obtained the gravitational wave fluxes to find how the parameters evolve through time.\\

\begin{figure}[h!]
\centering
\begin{subfigure}{0.8\textwidth}
   \includegraphics[width=\linewidth]{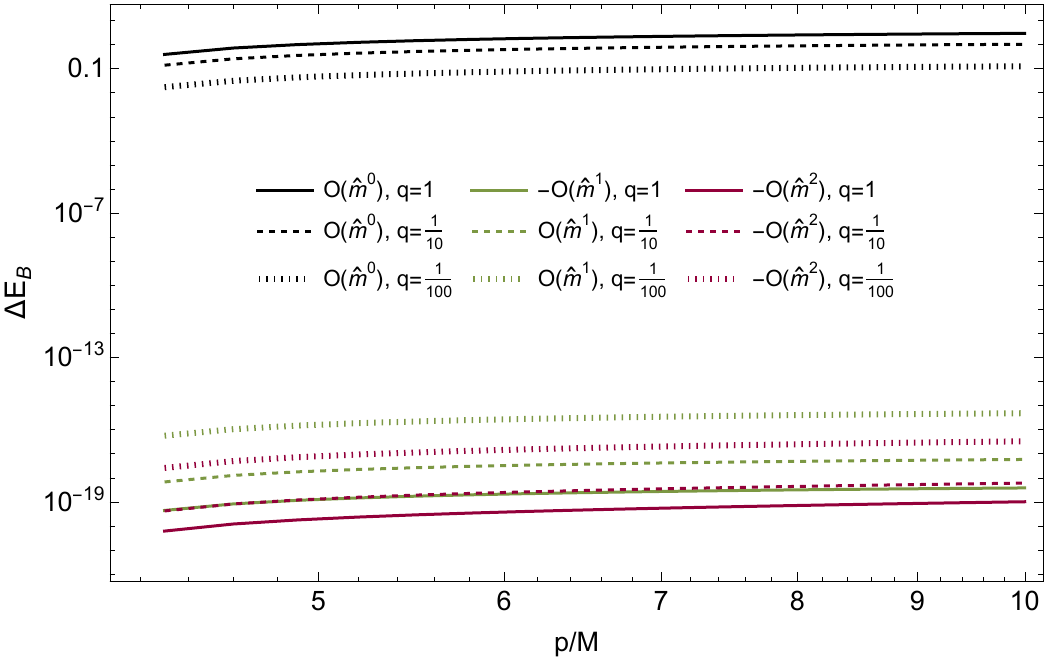}
       \centering
\end{subfigure}
\caption[]{The accumulated binding energy for eccentric orbits as function of $p$. For a system of $M=100 M_{\odot}$ the different mass ratios $q=1,\frac{1}{10},\frac{1}{100}$ are denoted with full, dashed and dotted curves respectively. Note the pink curves are multiplied with $-1$.}
\label{fig:Ebpp}
\end{figure}

\begin{figure}[h!]
\centering
\begin{subfigure}{0.8\textwidth}
   \includegraphics[width=\linewidth]{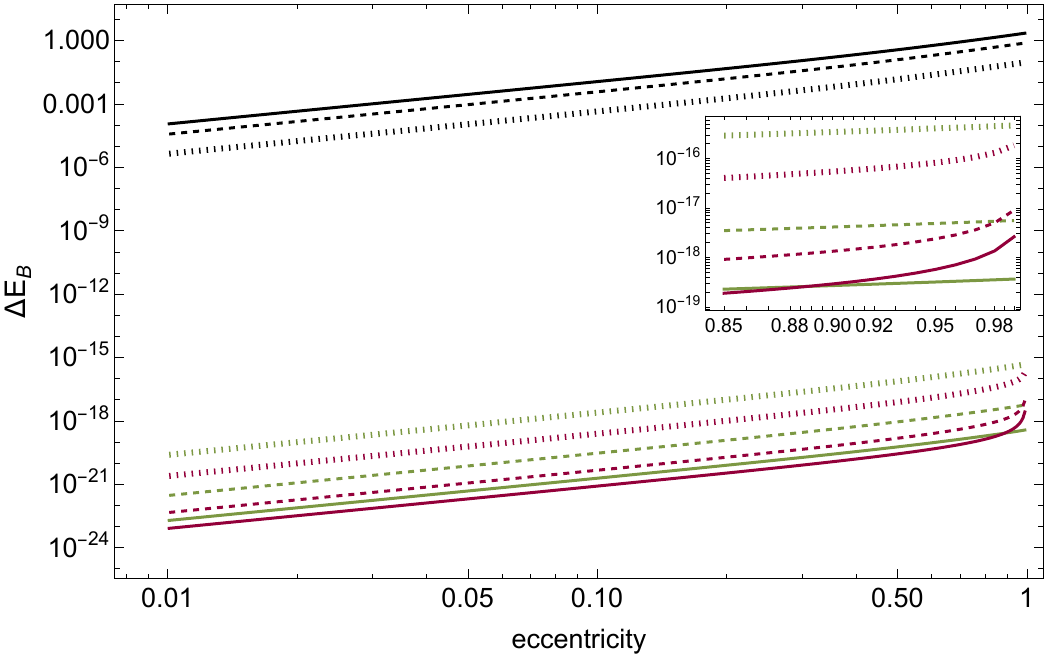}
       \centering
\end{subfigure}
\caption[]{The accumulated binding energy as function of the eccentricity. The legend of Fig.\ref{fig:Ebpp} is shared with this plot. The inset shows a zoom in of the linear and quadratic order contributions for high eccentricity.}
\label{fig:Ebeccee}
\end{figure}

\noindent The orbit average of a function f is given by
\be\label{eq:orbav}
<f> = \frac{\omega_{\xi}}{2\pi}\int_0^{2\pi}\frac{d\xi}{\mathcal{P}}f,
\ee
with $\omega_{\xi}$ the radial frequency
\be
\omega_{\xi} = \frac{2 \pi}{\int_0^{2\pi}\frac{d\xi}{\mathcal{P}}}.
\ee
It only makes sense to interpret quantities taking this orbital average. To plot the binding energy for eccentric orbits against the average separation $p$ we compute the orbital average of $E_B$ via \eqref{eq:orbav}. In Fig.~\ref{fig:Ebpp} for binary systems of $M=100 M_{\odot}$, $q=1,\frac{1}{10}, \frac{1}{100}$ and $\sqrt{\alpha}=1\textrm{km}$, $\tilde{m}=0.01$, $e=0.3$.
From~\ref{fig:Ebpp} we see similar behavior regarding the mass ratios as for the circular orbits. The signs of the accumulated curves are opposite, now the quadratic contribution has a negative sign. Note however that here we show the energy in terms of average separation and in the previous section we wrote the binding energy in terms of frequency. In the eccentric orbit case the allowed value of the scalar field mass for which the quadratic order contribution is still below the linear order contribution is raised as for this specific example can be seen as we chose $\hat{m}=0.01$ which for these black hole masses is above the maximum for circular orbits, see Fig.~\ref{fig:maxmhat}.\\
In Fig.~\ref{fig:Ebeccee} we show the accumulated binding energy as function of the eccentricity. Note that the legend of Fig.~\ref{fig:Ebpp} is shared with this figure. 
 From the figure we can see that for high values of the eccentricity around $e=1$, also shown more closely in the inset, the quadratic scalar mass contribution can become larger than the linear contribution. Hence the perturbative expansion breaks down for highly eccentric orbits. This can be explained in the following way. In the perturbative expansion the scalar mass contributions in the form of an exponential are Taylor expanded $$\frac{e^{-\mphi r}}{r}\sim \frac{1}{r}-\mphi +\frac{\mphi^2 r}{2} +\ldots.$$
This expansion is the same if we would expand in small $r$, hence also for large distances this perturbative expansion breaks down. For highly eccentric orbits the relative separation can become very large. It is interesting that in this regime of parameters space the scalar mass contributions are enhanced. Expansion to higher order in the scalar field mass or a non-perturbative treatment is required to study this regime further. 
\clearpage
\section{Conclusion and Outlook}\label{sec:conclusion}
We obtained the equations-of-motion, center-of-mass transformation and binding energy for circular and eccentric orbits for scalarized binary systems in massive sGB gravity up to first PN order and second order in the scalar field mass. We have shown explicitly the DIRE approach, solving the field integrals by splitting the integration domain, is still applicable on Klein-Gordon type equations and the final solution to 1PN does not depend on the near zone boundary $\mathcal{R}$. Furthermore, we found a constant far zone contribution which is not present in the massless scalar field limit at this PN order, though ultimately has no effect on observables. 
We allowed for arbitrary scalar masses for most of the calculations, however, to obtain explicit results required us to eventually limit to small scalar masses compared to the total mass of the binary. Specifically, in the near-zone calculations, we cut off the Taylor expansion in the field integral to second order, which effectively is similar to the small mass regime. For the far zone solution of the fields we did explicitly expand in the scalar mass, however this only gave a constant contribution which will not contribute to the radiation.

From the analysis of the binding energy as a function of frequency we identified the regime of validity of the small scalar mass expansion by requiring the correct hierarchy of the mass corrections, limiting the allowed scalar mass relative to the total mass of the binary. Within this scalar mass range and a coupling around the current constraint set by GW observations, we found the mass correction to the binding energy to be many orders of magnitude smaller than the point particle and massless scalar contributions. The mass contribution do increase for systems with more unequal masses for which the difference in scalar charges between the bodies is largest. They also increase for larger eccentricities, however for eccentricities close to the limit to unbound orbits, the perturbative expansion fails. This limit would be interesting to study further as the scalar mass contributions are enhanced, this requires higher orders in the perturbative expansion or an exact scalar mass treatment.

That the scalar mass corrections to the binding energy are many order of magnitude smaller does not necessarily mean they are out of observational interest in the context of GWs. First of all, during the inspiral small corrections can accumulate over the many GW cycles and become more relevant. Secondly, based on the results obtained for the inspiral gravitational and scalar losses for binary systems in massive scalar-tensor theory~\cite{Alsing:2011er,Diedrichs:2023foj} there is a non-perturbative resonance-like effect when multiples of the orbital frequencies match the scale associated to the scalar mass. The methods and results of this paper can serve as the foundation for computing these effects and their GW signatures.
Other follow ups on this work could be to extend to higher PN order and to include spinning black holes, other spin-orbit couplings and finite size effects to pave to way to more accurate waveform templates in modified gravity. Furthermore, higher order self interactions to the scalar field or massive scalar field extensions of other modified theories are interesting to consider. 
All in all our work obtained the higher curvature-scalar mass corrections to the scalarized binary dynamics for the first time and showed to extension of the DIRE and PN approach to solve the field equations to include Klein-Gordon type equations and the application of the Hadamard regularization within DIRE for a finite integration domain. These results represent an important step towards constructing  inspiral waveforms in massive sGB and can be directly used for this goal.  

\section{Acknowledgments}
We thank Robin Diedrichs for his usefull comments and effort in comparing the results. This publication is part of the Dutch Black Hole Consortium with project number NWA.1292.19.202 of the research programme NWA which is (partly) financed by the Dutch Research Council (NWO).
\clearpage

\appendix
\renewcommand{\theequation}{A.\arabic{equation}}
\setcounter{equation}{0}
\section{Regularized integration over a finite domain}\label{sec:appendixA}
Consider two open domains $\mathcal{D}_A$ and $\mathcal{D}_B$ that are disjoined $\mathcal{D}_A \bigcap \mathcal{D}_B=\emptyset$ and their union spans the near zone $\mathcal{M}$ bounded by radius $\mathcal{R}$; $\overline{\mathcal{D}_A \bigcup \mathcal{D}_B}=\mathcal{M}$. Furthermore $\mathbf{x}_A \in \mathcal{D}_A$ and $\mathbf{x}_B \in \mathcal{D}_B$ and two open balls around $\mathbf{x}_A$ and $\mathbf{x}_B$ being $\mathcal{B}_A$, $\mathcal{B}_B$ respectively. Now by definition, the partie-finie integral over $\mathcal{D}_B$ is given by
\begin{equation}
\operatorname{Pf} \int_{\mathcal{D}_B} d^3 \mathbf{x} F=\lim _{s \rightarrow 0}\left\{\int_{\mathcal{D}_B \backslash \mathcal{B}_B(s)} d^3 \mathbf{x} F+\sum_{b+3<0} \frac{s^{b+3}}{b+3} \int d \Omega_B f_B+\ln \left(\frac{s}{s_B}\right) \int d \Omega_B f_B\right\}
\end{equation}
We can rewrite the singular corrections rewriting the integral 
$$\sum_{b+3<0} \int_{\mathcal{M}\backslash \mathcal{B}_B} d^3x\, r_B^b f_{B,b}$$
with $r_B=|x-x_B|$. Now we like to shift the integration to $r_B$ via $\mathbf{x}=\mathbf{r_B}+\mathbf{x_B}$. The integration domain $|x|<\mathcal{R}$ then becomes $|r_B+x_B|<\mathcal{R}$. It is convenient to rewrite the integration over the simpler domain $|r_B|<\mathcal{R}$ plus corrections. Writing out the inequality $r_B^2 + 2 \mathbf{r_B}\cdot\mathbf{x_B} + x_B^2 < \mathcal{R}$ we can write 
$$
r_B<\mathcal{R}-x_B \cos \gamma_E+O\left(x_B^2 / \mathcal{R}\right),
$$
when $x_B / \mathcal{R} \ll 1$ which is the case in the NZ; here $\gamma_E$ is the angle between the vectors $\boldsymbol{r_B}$ and $\boldsymbol{x_B}$.

Then in spherical coordinates $(r_B, \theta, \phi)$ associated with the integral is

$$
\begin{aligned}
\int_{\mathcal{M}\backslash \mathcal{B}_B} d^3x\, r_B^b f_{B,b}=\int_{\mathcal{M}\backslash \mathcal{B}_B} d^3r_B\, r_B^b f_{B,b} &=\int d \Omega_B \int_s^{\mathcal{R}-x_B \cos \gamma_E+\cdots} r_B^{b+B} f_{B,b}  d r_B \\
& =\int d \Omega_B \int_s^{\mathcal{R}} r_B^{b+2} f_{B,b}  d r_B+\int d \Omega_B \int_{\mathcal{R}}^{\mathcal{R}-x_B \cos \gamma_E+\cdots} r_B^{b+2} f_{B,b}  d r_B
\end{aligned}
$$
The last integral can be written as
$$
\int(-x_B \cos \gamma_E) \mathcal{R}^2 \mathcal{R}^{b} f_{B,b} d \Omega_B=-\oint \mathcal{R}^{b} f_{B,b} \boldsymbol{x_B} \cdot \boldsymbol{d} \boldsymbol{S_B}
$$
to first order in $x_B / \mathcal{R}$; here, $d S_B^j:=\mathcal{R}^2 N_{r_B}^j d \Omega_B$, with $\boldsymbol{N_{r_B}}:=\boldsymbol{r_B} / r_B$. We thus obtain
\be
\bal
\sum_{b+3<0} \int_{\mathcal{M}\backslash \mathcal{B}_B} d^3x\, r_B^b f_{B,b} &= \int_{s}^{R} dr_B d\Omega_B r_B^{b+2} f_{B,b} - \int d\Omega_B\, r_B^{b} f_{B,b} \mathcal{R}^2 \mathbf{x_B}\cdot \mathbf{N_{rB}} + \mathcal{O}(x_B^2/\mathcal{R}) \\
&=\sum_{b+3<0}( \frac{\mathcal{R}^{b+3}}{b+3} \int d\Omega_B f_{B,b} - \frac{s^{b+3}}{b+3} \int d\Omega_B f_{B,b}\\
&- \int d\Omega_B\, r_B^{b+2} f_{B,b} \mathcal{R}^2 \mathbf{x_B}\cdot \mathbf{N_{rB}} + \mathcal{O}(x_B^2/\mathcal{R}) )
\eal
\ee
Hence

\be
\bal
\sum_{b+3<0} \frac{s^{b+3}}{b+3} \int d\Omega_B f_{B,b} &= -\sum_{b+3<0} \int_{\mathcal{M}\backslash \mathcal{B}_B} d^3x\, r_B^b f_{B,b} +\sum_{b+3<0}\frac{\mathcal{R}^{b+3}}{b+3} \int d\Omega_B f_{B,b}\\
&-\sum_{b+3<0}\int d\Omega_B\, r_B^{b+2} f_{B,b} \mathcal{R}^2 \mathbf{x_B}\cdot \mathbf{N_{rB}}
\eal
\ee
And for
\be
\bal
\int_{\mathcal{M}\backslash \mathcal{B}_B} d^3x\, r_B^{-3}f_B^{-3} &= \int d\Omega_B f_B^{-3} \int_{s}^{\mathcal{R}} dr_B r_B^{-1} - \int d\Omega_B f_B^{-3}r_B^{-3} \mathcal{R}^2\mathbf{x_B}\cdot\mathbf{N_{r_B}}\\
&= (\ln(\frac{\mathcal{R}}{s_B})-\ln(\frac{s}{s_B}))\int d\Omega_B f_B^{-3}  - \int d\Omega_B f_B^{-3}r_B^{-3} \mathcal{R}^2\mathbf{x_B}\cdot\mathbf{N_{r_B}}
\eal
\ee
So we can write
\be
\ln \left(\frac{s}{s_B}\right) \int d \Omega_B f_B = -\int_{\mathcal{M}\backslash \mathcal{B}_B} d^3x\, r_B^{-3}f_B^{-3} +\ln(\frac{\mathcal{R}}{s_B})\int d\Omega_B f_B^{-3}  - \int d\Omega_B f_B^{-3}r_B^{-3} \mathcal{R}^2\mathbf{x_B}\cdot\mathbf{N_{r_B}}
\ee
Substituting these expressions in the PF integral over $\mathcal{D}_B$

\begin{equation}
\bal
\operatorname{Pf} \int_{\mathcal{D}_B} d^3 \mathbf{x} F &=\lim _{s \rightarrow 0}\left\{\int_{\mathcal{D}_B \backslash \mathcal{B}_B(s)} d^3 \mathbf{x} F-\sum_{b+3\leq0} \underbrace{\int_{\mathcal{M}\backslash \mathcal{B}_B}}_{\mathcal{D}_B\backslash\mathcal{B}_B + \mathcal{D}_A} d^3x\, r_B^b f_{B,b} +\sum_{b+30}\frac{\mathcal{R}^{b+3}}{b+3} \int d\Omega_B f_{B,b}\right.\\
&\left.+\ln(\frac{\mathcal{R}}{s_B})\int d\Omega_B f_B^{-3} -\sum_{b+3\leq0}\int d\Omega_B\, r_B^{b} f_{B,b} \mathcal{R}^2 \mathbf{x_B}\cdot \mathbf{N_{rB}}\right\}\\
&= \lim _{s \rightarrow 0}\left\{\int_{\mathcal{D}_B} d^3 \mathbf{x} \tilde{F}_B-\sum_{b+3\leq0}\int_{\mathcal{D}_A} d^3x\, r_B^b f_{B,b} +\sum_{b+30}\frac{\mathcal{R}^{b+3}}{b+3} \int d\Omega_B f_{B,b}\right.\\
&\left.+\ln(\frac{\mathcal{R}}{s_B})\int d\Omega_B f_B^{-3} -\sum_{b+3\leq0}\int d\Omega_B\, r_B^{b} f_{B,b} \mathcal{R}^2 \mathbf{x_B}\cdot \mathbf{N_{rB}}\right\} 
\eal
\end{equation}
With
\be
\tilde{F}_B = F-\sum_{b+3\leq 0} r_B^b f_{B,b}.
\ee
Similarly the PF integral over $\mathcal{D}_A$ can be written as
\begin{equation}
\operatorname{Pf} \int_{\mathcal{D}_A} d^3 \mathbf{x} F=\lim _{s \rightarrow 0}\left\{\int_{\mathcal{D}_A \backslash \mathcal{B}_A(s)} d^3 \mathbf{x} F+\sum_{b+3<0} \frac{s^{b+3}}{b+3} \int d \Omega_A f_A+\ln \left(\frac{s}{s_A}\right) \int d \Omega_A f_A\right\}
\end{equation}
In the total integral over $\mathcal{D}_A+\mathcal{D}_B$ the left over $-\sum_{b+3\leq0}\int_{\mathcal{D}_A} d^3x\, r_B^b f_{B,b}$ in the integration over $\mathcal{D}_B$ can be combined with the integral over $F$ in $\mathcal{D}_A$ to have the integration again over $\tilde{F}_B$. So in total we have
\begin{equation}
\bal
\operatorname{Pf} \int d^3 \mathbf{x} F&=\lim _{s \rightarrow 0}\left\{\int_{\mathcal{M}\backslash\mathcal{B}_A} d^3 \mathbf{x} \tilde{F}_B+\sum_{b+3<0} \frac{s^{b+3}}{b+3} \int d \Omega_A f_A+\ln \left(\frac{s}{s_A}\right) \int d \Omega_A f_A\right.\\
&\left. +\sum_{b+30}\frac{\mathcal{R}^{b+3}}{b+3} \int d\Omega_B f_{B,b}+\ln(\frac{\mathcal{R}}{s_B})\int d\Omega_B f_B^{-3} -\sum_{b+3\leq0}\int d\Omega_B\, r_B^{b} f_{B,b} \mathcal{R}^2 \mathbf{x_B}\cdot \mathbf{N_{rB}}\right\}\\
&=\lim _{s \rightarrow 0}\left\{\int d\Omega_A \int_{s}^{\mathcal{R}} d r_A \tilde{F}_B r_A^2 - \int d\Omega_A \tilde{F}_B \mathcal{R}^2 \mathbf{x}_A\cdot \mathbf{N_{r_A}}\right.\\
&\left. +\sum_{b+3<0} \frac{s^{b+3}}{b+3} \int d \Omega_A f_A+\ln \left(\frac{s}{s_A}\right) \int d \Omega_A f_A\right.\\
&\left. +\sum_{b+3<0}\frac{\mathcal{R}^{b+3}}{b+3} \int d\Omega_B f_{B,b}+\ln(\frac{\mathcal{R}}{s_B})\int d\Omega_B f_B^{-3} -\sum_{b+3\leq0}\int d\Omega_B\, r_B^{b} f_{B,b} \mathcal{R}^2 \mathbf{x_B}\cdot \mathbf{N_{rB}}\right\}
\eal
\end{equation}

\renewcommand{\theequation}{B.\arabic{equation}}
\setcounter{equation}{0}
\section{Coefficients of the field and dynamics expressions}\label{sec:AppB}
We moves the explicit expressions of the caligraphic coefficients of the field solutions in Sec.~\ref{sec:1PNNZfields} and binary dynamics expressions in Sec.~\ref{sec:BinaryDyn} to this appendix to keep the main text readable. We list the expressions here below and link to the equations in the main text. 

Coefficients of \eqref{eq:NZsolh00}
\be\label{eq:Coeffh00NZ}
\bal
\mathcal{A}_{h^{00}}&=2\, v_A^i v_{Ai} + 2\, v_A^i v_{Bi}
    - 2\, n_{AB}^i n_{AB}^j v_{Ai} v_{Bj},\\
\mathcal{B}_{h^{00}}&= 7\,  - 2\,  \frac{m_B}{m_A}
    -  \alpha_A^2 - 2\,  \frac{m_B}{m_A} \, \alpha_A \alpha_B, \\
\mathcal{C}_{h^{00}}&= 4 (\alpha_A + \frac{m_B}{m_A} \left( \alpha_A + \alpha_B \right)  ),\\
\mathcal{D}_{h^{00}}&=
     \alpha_A^2 + \frac{m_B}{m_A} \, \alpha_A \alpha_B,\\
\mathcal{E}_{h^{00}}&=
     \alpha_A^2 + 2\, \frac{m_B}{m_A} \, \alpha_A \alpha_B, \\
\mathcal{F}_{h^{00}}&=
    - 4\,  \alpha_A
    + \frac{9}{4}\, \frac{m_B}{m_A} \, \alpha_A
    + \frac{4}{3}\, \frac{m_B}{m_A} \, \alpha_B.
\eal
\ee

Coefficients of \eqref{eq:NZsolhii}
\be\label{eq:CoeffhiiNZ}
\bal
\mathcal{A}_{h_i^i}&=
    m_A^2 - 2\, m_A m_B
    + m_A^2 \alpha_A^2 - 2\, m_A m_B\, \alpha_A \alpha_B,\\
\mathcal{B}_{h_i^i}&=m_A^2 \alpha_A^2 - m_A m_B\, \alpha_A \alpha_B,\\
\mathcal{C}_{h_i^i}&=
    m_A^2 \alpha_A^2 + 2\, m_A m_B\, \alpha_A \alpha_B.
\eal
\ee
Coefficients of \eqref{eq:NZsolphi}
\be\label{eq:CoeffphiNZ}
\bal
\mathcal{A}_{\delta\varphi}&= \alpha_A + \alpha_A^2 \alpha_B + \alpha_B \beta_A,\\
\mathcal{B}_{\delta\varphi}&=
    v_{Ai} v_A^i - v_{Ai} v_B^i
    + n_{AB}^i n_{AB}^j v_{Ai} v_{Bj},\\
\mathcal{C}_{\delta\varphi}&= m_A^2 + 2\, m_A m_B,\\
\mathcal{D}_{\delta\varphi}&= 
    m_B \alpha_A
    + m_A \, \alpha_A^3
    + m_B \, \alpha_B
    + 2\, m_B \alpha_A^2 \alpha_B
    + m_B \, \alpha_A \alpha_B^2
    + m_A \, \alpha_A \beta_A
    + 2\, m_B \alpha_B \beta_A
    + m_B \, \alpha_A \beta_B,\\
\mathcal{E}_{\delta\varphi}&= 
    v_{Ai} v_A^i - v_{Ai} v_B^i,\\
\mathcal{F}_{\delta\varphi}&=
    2\, m_A \alpha_A + \frac{5}{2}\, m_B \alpha_A
    + m_A \, \alpha_A^3
    + 2\, m_B \alpha_B
    + 2\, m_B \alpha_A^2 \alpha_B
    + m_B \, \alpha_A \alpha_B^2
    + \frac{m_B^2}{m_A} \alpha_B^3\\
    &+ m_A \, \alpha_A \beta_A
    + 2\, m_B \alpha_B \beta_A
    + m_B \, \alpha_A \beta_B
    + \frac{m_B^2}{m_A} \alpha_B \beta_B,\\
\mathcal{G}_{\delta\varphi}&= - m_A^2 \alpha_A - m_A m_B \left(\alpha_A + \alpha_B\right),\\
\mathcal{H}_{\delta\varphi}&=v_{Ai} v_A^i - v_{Ai} v_B^i- n_{AB}^i n_{AB}^j v_{Ai} v_{Bj}.
\eal
\ee

Coefficients of \eqref{eq:L10}
\be\label{eq:CoeffL10}
\bal
\mathcal{A}_{\mathcal{L}^{(10)}}&=
    \frac{3}{2}\, v_{Ai} v_A^i
    + \frac{3}{2}\, v_{Bi} v_B^i
    - \frac{7}{2} v_{Ai} v_B^i
    - \frac{1}{2}\, n_{AB}^i n_{AB}^j v_{Ai} v_{Bj},\\
\mathcal{B}_{\mathcal{L}^{(10)}}&=
    - \frac{1}{2}\, v_{Ai} v_A^i
    - \frac{1}{2}\, v_{Bi} v_B^i
    + \frac{1}{2}\, v_{Ai} v_B^i
    - \frac{1}{2}\, n_{AB}^i n_{AB}^j v_{Ai} v_{Bj},\\
\mathcal{C}_{\mathcal{L}^{(10)}}&=\frac{1}{2}(1 + 2\alpha_A\alpha_B +\alpha_A^2\alpha_B^2),\\
\mathcal{D}_{\mathcal{L}^{(10)}}&=
    m_B\, \alpha_B^2 \beta_A + m_A\, \alpha_A^2 \beta_B,\\
\mathcal{E}_{\mathcal{L}^{(10)}}&=
     m_A m_B \bigl(2\, m_A + m_B\bigr) \alpha_A
    + m_A m_B \bigl(m_A + 2\, m_B\bigr) \alpha_B.
\eal
\ee
Coefficients of \eqref{eq:L11}
\be\label{eq:CoeffL11}
\bal
\mathcal{A}_{\mathcal{L}^{(11)}}&=
    m_A^2 \alpha_A^2\, v_{Ai} v_A^i
    + m_B^2 \alpha_B^2\, v_{Bi} v_B^i\\
\mathcal{B}_{\mathcal{L}^{(11)}}&=
     v_{Ai} v_A^i + v_{Bi} v_B^i
    - \, v_{Ai} v_B^i,\\
\mathcal{C}_{\mathcal{L}^{(11)}}&= m_A \alpha_A^2 + m_B \alpha_B^2,\\
\mathcal{D}_{\mathcal{L}^{(11)}}&=m_A + m_B + m_A \beta_A + m_B \beta_B,\\
\mathcal{E}_{\mathcal{L}^{(11)}}&=m_A \alpha_A^3 \alpha_B + m_B \alpha_A \alpha_B^3,\\
\mathcal{F}_{\mathcal{L}^{(11)}}&=m_B \alpha_B^2 \beta_A + m_A \alpha_A^2 \beta_B.\\
\eal
\ee
Coefficients of \eqref{eq:L12}
\be\label{eq:CoeffL12}
\bal
\mathcal{A}_{\mathcal{L}^{(21)}} &=
    - \, v_{Ai} v_A^i
    - \, v_{Bi} v_B^i
    + \, v_{Ai} v_B^i
    + \, n_{AB}^i n_{AB}^j v_{Ai} v_{Bj},\\
\mathcal{B}_{\mathcal{L}^{(21)}}&= -m_A \alpha_A^2 -\frac{m_A^2\alpha_A^4}{2m_B} -m_A\alpha_A^3 \alpha_B -m_B\alpha_B^2 -m_A\alpha_A^2\alpha_B^2 -m_B\alpha_A^2 \alpha_B^2\\
&- m_B\alpha_A\alpha_B^3-\frac{m_B^2\alpha_B^4}{2m_A} -\frac{m_A^2\alpha_A^2\beta_A}{2m_B} -m_B\alpha_B^2\beta_A -m_A\alpha_A^2\beta_B-\frac{m_B^2\alpha_B^2\beta_B}{2m_A},\\
\mathcal{C}_{\mathcal{L}^{(21)}} &=\frac{5}{2}\,(m_A + m_B)
    + m_A \beta_A + m_B \beta_B,\\
\mathcal{D}_{\mathcal{L}^{(21)}} &=
    m_A m_B \bigl(m_A \alpha_A^2 + m_B \alpha_B^2\bigr)
    + 2\, m_A m_B (m_A + m_B)\, \alpha_A \alpha_B,\\
\mathcal{E}_{\mathcal{L}^{(21)}} &=
    \Bigl(-\frac{4}{3} m_A^2 m_B + \frac{9}{32} m_A m_B^2\Bigr) \alpha_A
    + \Bigl(\frac{9}{32} m_A^2 m_B - \frac{4}{3} m_A m_B^2\Bigr) \alpha_B.\\
\eal
\ee
Coefficients of \eqref{eq:a10}
\be\label{eq:Coeffarel10}
\bal
\mathcal{A}_{a_{rel}^{(10)}}&= 4\bigl(m_A^2 + m_B^2\bigr) + 10\, m_A m_B,\\
\mathcal{B}_{a_{rel}^{(10)}}&= 4\bigl(m_A^2 + m_B^2\bigr) + 12\, m_A m_B,\\
\mathcal{C}_{a_{rel}^{(10)}}&= m_B \alpha_B^2 \beta_A + m_A \alpha_A^2 \beta_B,\\
\mathcal{D}_{a_{rel}^{(10)}}&= 
    \bigl(8\, m_A^2 + 12\, m_A m_B + 4\, m_B^2\bigr)\alpha_A
    + \bigl(4\, m_A^2 + 12\, m_A m_B + 8\, m_B^2\bigr)\alpha_B,\\
\mathcal{E}_{a_{rel}^{(10)}}&= m_A^3 + 6\, m_A^2 m_B + 6\, m_A m_B^2 + m_B^3.\\
\eal
\ee
Coefficients of \eqref{eq:a11}
\be\label{eq:Coeffarel11}
\bal
\mathcal{A}_{a_{rel}^{(11)}}&= 
    m_A\bigl(m_A + \tfrac{1}{2} m_B\bigr)\,  \alpha_A^2
    + m_B\bigl(m_B + \tfrac{1}{2} m_A\bigr)\, \alpha_B^2,\\
\mathcal{B}_{a_{rel}^{(11)}}&= 
    m_A\bigl(m_A + \tfrac{1}{2} m_B\bigr)\, \alpha_A^3 \alpha_B
    + m_B\bigl(m_B + \tfrac{1}{2} m_A\bigr)\, \alpha_A \alpha_B^3,\\
\mathcal{C}_{a_{rel}^{(11)}}&=
    m_A(m_A + m_B)\, \alpha_A \alpha_B \beta_A
    + m_B(m_A + m_B)\, \alpha_A \alpha_B \beta_B,\\
\mathcal{D}_{a_{rel}^{(11)}}&= 
    m_B(m_A + m_B)\, \alpha_B^2 \beta_A
    + m_A(m_A + m_B)\, \alpha_A^2 \beta_B.
\eal
\ee
Coefficients of \eqref{eq:a12}
\be\label{eq:Coeffarel12}
\bal
\mathcal{A}_{a_{rel}^{(12)}}&= m_A(m_A + m_B)\,\alpha_A^2
    + m_B(m_A + m_B)\,\alpha_B^2
    - m_A m_B\,\alpha_A \alpha_B
    - m_A m_B\,\alpha_A^2 \alpha_B^2,\\
\mathcal{B}_{a_{rel}^{(12)}}&= 
    \Bigl(\frac{8}{3} m_A^2 + \frac{101}{48} m_A m_B
    - \frac{9}{16} m_B^2\Bigr)\alpha_A
    + \Bigl(-\frac{9}{16} m_A^2 + \frac{101}{48} m_A m_B
    + \frac{8}{3} m_B^2\Bigr)\alpha_B.
\eal
\ee
Coefficients of \eqref{eq:r10}
\be\label{eq:Coeffr10}
\bal
\mathcal{A}_{r^{(10)}}&= -m_A^2 -m_B^2 - \frac{5}{3} m_A m_B -\frac{4}{3}(m_A^2+m_B^2) \alpha_A \alpha_B - 2 m_A m_B \alpha_A\alpha_B \\
    &-\frac{1}{3}(m_A^2 +m_A m_B +m_B^2)\alpha_A^2\alpha_B^2 -\frac{1}{3}m_A m_B \alpha_B^2 \beta_A -\frac{1}{3} m_B^2 \alpha_B^2 \beta_A - \frac{1}{3} m_A^2 \alpha_A^2 \beta_B\\\nonumber
    &-\frac{1}{3}m_A m_B \alpha_A^2 \beta_B,\\
\mathcal{B}_{r^{(10)}}&= 
    m_A^2 + 2 m_A m_B + m_B^2 + 2 m_A^2 \alpha_A \alpha_B + 4 m_A m_B \alpha_A \alpha_B+ 2 m_B^2 \alpha_A \alpha_B + m_A^2 \alpha_A^2 \alpha_B^2\\
    &+2 m_A m_B \alpha_A^2 \alpha_B^2 + m_B^2 \alpha_A^2\alpha_B^2,\\
\mathcal{C}_{r^{(10)}}&=2m_A\alpha_A + m_B \alpha_A + m_A\alpha_B +2 m_B\alpha_B.
\eal
\ee
Coefficients of \eqref{eq:r11}
\be\label{eq:Coeffr11}
\bal
\mathcal{A}_{r^{(11)}}&=m_A^2 \alpha_A^2 + \frac{1}{2}m_A m_B \alpha_A^2 + m_A m_B \alpha_A \alpha_B + m_A^2 \alpha_A^3 \alpha_B + \frac{1}{2} m_A m_B \alpha_A^3 \alpha_B \\
&+ \frac{1}{2}m_A m_B \alpha_B^2 + m_B^2 \alpha_B^2 +m_A m_B \alpha_A^2 \alpha_B^2 +\frac{1}{2} m_A m_B \alpha_A \alpha_B^3 + m_B^2 \alpha_A \alpha_B^3\\
&+ m_A^2 \alpha_A \alpha_B \beta_A + m_A m_B \alpha_A \alpha_B \beta_A + m_A m_B \alpha_B^2 \beta_A +m_B^2 \alpha_B^2\beta_A\\
&+ m_A^2 \alpha_A^2 \beta_B + m_A m_B \alpha_A^2 \beta_B + m_A m_B \alpha_A \alpha_B \beta_B +m_B^2 \alpha_A \alpha_B\beta_B.
\eal
\ee
Coefficients of \eqref{eq:r12}
\be\label{eq:Coeffr12}
\bal
\mathcal{A}_{r^{(12)}}&= -\frac{1}{6} m_A^3 \alpha_A \alpha_B - \frac{1}{2} m_A^2 m_B \alpha_A \alpha_B -\frac{1}{2} m_A m_B^2\alpha_A \alpha_B -\frac{1}{6}m_B^3 \alpha_A \alpha_B - \frac{1}{3}m_A^3 \alpha_A^2 \alpha_B^2\\
&- m_A^2 m_B \alpha_A^2 \alpha_B^2 -m_A m_B^2 \alpha_A^2 \alpha_B^2 -\frac{1}{3}m_B^3 \alpha_A^2\alpha_B^2-\frac{1}{6}m_A^3\alpha_A^3\alpha_B^3\\
&-\frac{1}{2}m_A^2 m_B \alpha_A^3\alpha_B^3 -\frac{1}{2}m_A m_B^2\alpha_A^3\alpha_B^3 -\frac{1}{6}m_B^3\alpha_A^3\alpha_B^3,\\
\mathcal{B}_{r^{(12)}}&= 
   -\frac{1}{3} m_A^3\alpha_A^2 -\frac{2}{3}m_A^2 m_B\alpha_A^2- \frac{1}{9}m_A^2 m_B\alpha_A^3\alpha_B\beta_B -\frac{1}{18}m_A m_B^2\alpha_A^3\alpha_B\beta_B\\
&-\frac{1}{3}m_A m_B^2\alpha_A^2 -\frac{1}{6}m_A^3 \alpha_A \alpha_B -\frac{10}{9}m_A^2m_B\alpha_A\alpha_B -\frac{10}{9}m_A m_B^2\alpha_A\alpha_B-\frac{1}{6}m_B^3\alpha_A\alpha_B\\
&-\frac{1}{3}m_A^3\alpha_A^3\alpha_B - \frac{2}{3}m_A^2 m_B \alpha_A^3\alpha_B -\frac{1}{3}m_A m_B^2\alpha_A^3\alpha_B -\frac{1}{3}m_A^2 m_B\alpha_B^2 - \frac{2}{3}m_A m_B^2\alpha_B^2-\frac{m_B^3\alpha_B^2}{3}\\
&+\frac{1}{9}m_A^3\alpha_A^2\alpha_B^2-\frac{8}{9}m_A^2 m_B \alpha_A^2\alpha_B^2-\frac{8}{9}m_A m_B^2\alpha_A^2\alpha_B^2+\frac{1}{9}m_B^3\alpha_A^2\alpha_B^2 -\frac{1}{3}m_A^2 m_B\alpha_A\alpha_B^3\\
&-\frac{2}{3}m_A m_B^2 \alpha_A\alpha_B^3 -\frac{1}{3}m_B^3\alpha_A\alpha_B^3+\frac{5}{18}m_A^3 \alpha_A^3\alpha_B^3 +\frac{2}{9}m_A^2 m_B\alpha_A^3\alpha_B^3 +\frac{2}{9}m_A m_B^2\alpha_A^3\alpha_B^3\\
&+\frac{5}{18}m_B^3 \alpha_A^3\alpha_B^3-\frac{1}{18}m_A^2 m_B \alpha_A\alpha_B^3\beta_A -\frac{1}{9}m_A m_B^2\alpha_A \alpha_B^3\beta_A -\frac{1}{18}m_B^3\alpha_A \alpha_B^3 \beta_A -\frac{1}{18}m_A^3\alpha_A^3\alpha_B\beta_B,\\
\mathcal{C}_{r^{(12)}}&=\frac{8}{9}m_A\alpha_A-\frac{3}{16}m_B \alpha_A - \frac{3}{16}m_A \alpha_B +\frac{8}{9}m_B \alpha_B -\frac{4}{9}m_A\alpha_A^2\alpha_B -\frac{41}{48}m_B\alpha_A^2\alpha_B -\frac{41}{48}m_A\alpha_A\alpha_B^2-\frac{4}{9}m_B\alpha_A\alpha_B^2.
\eal
\ee
Coefficients of \eqref{eq:E10}
\be\label{eq:CoeffE10}
\bal
\mathcal{A}_{E^{(10)}}&=(m_A^2+m_B^2)(1+\alpha_A\alpha_B)\bigl(-9+7\,\alpha_A\alpha_B\bigr)\nonumber\\
      &
      + m_A m_B\bigl(-19 - 6\,\alpha_A\alpha_B + 13\,\alpha_A^2\alpha_B^2\bigr)
     + 4\,M\bigl(m_B\,\alpha_B^2\beta_A + m_A\,\alpha_A^2\beta_B\bigr),\\
\mathcal{B}_{E^{(10)}}&=m_A(2\alpha_A+\alpha_B) + m_B(\alpha_A+2\alpha_B).
\eal
\ee
Coefficients of \eqref{eq:E11}
\be\label{eq:CoeffE11}
\bal
\mathcal{A}_{E^{(11)}}&=-4 \alpha_A^3 \alpha_B m_A^3 m_B-2 \alpha_A^2 m_A^3 m_B \left(3 \alpha_B^2+2 \beta_B+2\right)\\
&-2
   \alpha_A \alpha_B (2 \beta_A+3) m_A^3 m_B-5 \alpha_A^3 \alpha_B m_A^2 m_B^2-10 \alpha_A^2 \alpha_B^2 m_A^2
   m_B^2-4 \alpha_A^2 \beta_B m_A^2 m_B^2-5 \alpha_A^2 m_A^2 m_B^2\\
   &-\alpha_A \alpha_B m_A^2 m_B^2 \left(5 \alpha_B^2+4 \beta_A+4 \beta_B+10\right)-4 \alpha_B^2 \beta_A m_A^2 m_B^2-5 \alpha_B^2 m_A^2 m_B^2-6 \alpha_A^2 \alpha_B^2 m_A m_B^3\\
   &-2 \alpha_A \alpha_B m_A m_B^3 \left(2 \alpha_B^2+2 \beta_B+3\right)-4 \alpha_B^2 \beta_A m_A
   m_B^3-4 \alpha_B^2 m_A m_B^3,\\
\mathcal{B}_{E^{(11)}}&=6 \alpha_A^2
   m_A^4 (\alpha_A \alpha_B+1)+12 \alpha_A^3 \alpha_B m_A^3 m_B+12 \alpha_A^2 \left(\alpha_B^2+1\right) m_A^3 m_B\\
   &+12
   \alpha_A \alpha_B m_A^3 m_B+6 \alpha_A^3 \alpha_B m_A^2 m_B^2+24 \alpha_A^2 \alpha_B^2 m_A^2 m_B^2+6
   \alpha_A^2 m_A^2 m_B^2+6 \alpha_A \alpha_B \left(\alpha_B^2+4\right) m_A^2 m_B^2\\
   &+6 \alpha_B^2 m_A^2 m_B^2+12
   \alpha_A^2 \alpha_B^2 m_A m_B^3+12 \alpha_A \alpha_B \left(\alpha_B^2+1\right) m_A m_B^3+12 \alpha_B^2 m_A
   m_B^3+6 \alpha_B^2 m_B^4 (\alpha_A \alpha_B+1).
\eal
\ee
Coefficients of \eqref{eq:E12}
\be\label{eq:CoeffE12}
\bal
\mathcal{A}_{E^{(12)}}&=\alpha_A \left(91 \alpha_B^2-128\right) m_A+27 \alpha_B
   m_A+\alpha_A m_B (91 \alpha_A \alpha_B+27)-128 \alpha_B m_B,\\
\mathcal{B}_{E^{(12)}}&=-240 \alpha_A \alpha_B m_A^3 m_B (\alpha_A \alpha_B+1)^2 (m_A+m_B)(\alpha_A
   \alpha_B G+G)^2\\
   &-480 \alpha_A \alpha_B m_A^2 m_B^2 (\alpha_A \alpha_B+1)^2 (m_A+m_B) (\alpha_A \alpha_B G+G)^2-240 \alpha_A^3 \alpha_B^3 m_A m_B^3 (m_A+m_B) (\alpha_A \alpha_B G+G)^2\\
   &-480 \alpha_A^2\alpha_B^2 m_A m_B^3 (m_A+m_B) (\alpha_A \alpha_B G+G)^2
   -240 \alpha_A \alpha_B m_A m_B^3(m_A+m_B) (\alpha_A \alpha_B G+G)^2,\\
\mathcal{C}_{E^{(12)}}&=96 (\alpha_A \alpha_B+1) (m_A+m_B)^2 (\alpha_A \alpha_B G+G)^2\\
   &\left(m_A m_B (\alpha_A m_A (\alpha_A+2 \alpha_B)+\alpha_B m_B (2 \alpha_A+\alpha_B)) \left(6 \log (c)+3 \left(\log (\alpha_A \alpha_B+1)\right.\right.\right.\\
   &\left.\left.\left.-\log \left(\sqrt[3]{G}
   (\alpha_A \alpha_B+1)^2 (m_A+m_B)\right)+\log (x)\right)-2 \log (G)\right)+3 \log (\Lambda ) (m_A+m_B) (\alpha_A m_A+\alpha_B
   m_B)^2\right)\\
   &+144 \alpha_A^2 m_A^4 (\alpha_A \alpha_B+1) \left(\alpha_A^2+\beta_A\right) (m_A+m_B) (\alpha_A \alpha_B G+G)^2\\
   &+24 \alpha_A m_A^3 m_B (\alpha_A \alpha_B+1) (m_A+m_B) \left(\alpha_A \left(6 \alpha_A^2+12 \alpha_A \alpha_B+17 \alpha_B^2+4\right)\right.\\
   &\left.+6 \beta_A (\alpha_A+2 \alpha_B)+23 \alpha_B\right) (\alpha_A \alpha_B
   G+G)^2+48 \alpha_A^2 \beta_B m_A^3 m_B (7 \alpha_A \alpha_B+6) (m_A+m_B) (\alpha_A \alpha_B
   G+G)^2\\
   &+24 m_A^2 m_B^2 (m_A+m_B) (\alpha_A \alpha_B G+G)^2 \left(12 \alpha_A^4 \alpha_B^2+\alpha_A^3
   \alpha_B \left(29 \alpha_B^2+14 \beta_B+16\right)\right.\\
   &\left.+2 \alpha_A^2 \left(\alpha_B^2 \left(6 \alpha_B^2+6 \beta_A+6 \beta_B+35\right)+6
   \beta_B+2\right)+\alpha_A \alpha_B \left(2 \alpha_B^2 (7 \beta_A+8)+12 \beta_A+12 \beta_B+41\right)\right.\\
   &\left.+4 \alpha_B^2 (3 \beta_A+1)\right)+144 \alpha_B^2 m_B^4 (\alpha_A \alpha_B+1) \left(\alpha_B^2+\beta_B\right) (m_A+m_B) (\alpha_A \alpha_B
   G+G)^2\\
   &+408 \alpha_A^3 \alpha_B^3 m_A m_B^3 (m_A+m_B) (\alpha_A \alpha_B G+G)^2\\
   &+96 \alpha_A^2
   \alpha_B^2 m_A m_B^3 \left(3 \alpha_B^2+3 \beta_B+10\right) (m_A+m_B) (\alpha_A \alpha_B G+G)^2\\
   &+48 \alpha_B^2 m_A m_B^3 (m_A+m_B) \left(3 \alpha_B^2+6 \beta_A+3 \beta_B+2\right) (\alpha_A \alpha_B G+G)^2\\
   &+24 \alpha_A \alpha_B m_A m_B^3 (m_A+m_B) \left(2 \alpha_B^2 \left(3 \alpha_B^2+7 \beta_A+3 \beta_B+8\right)+12 \beta_B+23\right)
   (\alpha_A \alpha_B G+G)^2.
\eal
\ee

\renewcommand{\theequation}{C.\arabic{equation}}
\setcounter{equation}{0}
\section{Near zone fields evaluated at Far zone field point}\label{sec:AppC}
The source terms in the far zone only contain field dependent terms (the binary system only lies in the near zone). So we need to obtain the solution to the fields integrated over the NZ and evaluated at $x$ in the far zone as well as they act as a source. To first order in the weak field expansion in $G$ the total field solutions only have a near zone contribution as the far zone contribution is sourced by the fields itself so can only come in at the second iteration. Hence in this section we compute the near zone fields evaluated in the far zone $x$. As only $h^{00}$ and $\delta\varphi$ show up as source terms in \eqref{eq:PNsources} we only focus on those solutions.\\

\noindent We'll begin with the scalar field equation, starting from \eqref{eq:FIvarphiFZ}. When the field point is in the far zone and the integration over the near zone, we have $|x|>\mathcal{R}$ and $|x'|\ll |x|$. We can therefore Taylor expand around $\mathbf{x}' = 0$. As the dependence on $\mathbf{x}'$ always come as the distance $|x-x'|$ one can replace the derivatives  $\partial_{\mathbf{x}'}$ to $-\partial_{\mathbf{x}}$ in the Taylor expansion.
\be\label{eq:phiNZFZtaylor}
\bal
\varphi &= -\frac{G}{c^4}\left[\sum_{\ell=0}^{\infty}\frac{(-1)^{\ell}}{\ell!} \partial_L \left(\frac{1}{|x|} \int_{\mathcal{M}} \tilde{\mu}_s(ct-|x|,x') x'^L d^3x'\right)\right.\\
&\left.-\frac{1}{2}\mphi^2 \sum_{\ell=0}^{\infty}\frac{(-1)^{\ell}}{\ell!}  \left( \int_{\mathcal{M}}d^3x' x'^L\int_{-\infty}^{0} dt' \partial_L(\Theta(t-t' -|x|/c)) \tilde{\mu}_s(t',x') \right) \right].
\eal
\ee

\noindent For the second line in \eqref{eq:phiNZFZtaylor} the derivatives are only taken over the $\Theta$ function as the source term does not depend on $\mathbf{x}$. For this we find the following
\be
\bal
\frac{\partial}{\partial x^i}\Theta(\underbrace{t-t'-|x|/c}_{\tau(|x|)}) &= \frac{\partial}{\partial \tau} \Theta(\tau) \frac{\partial \tau}{\partial x^i}\\
&=\delta(t-t'-|x|/c) \frac{x^i}{|x|},
\eal
\ee
and take the higher order derivatives outside the integrals
\be\label{eq:phiNZFZtaylor}
\bal
\varphi &= -\frac{G}{c^4}\left[\sum_{\ell=0}^{\infty}\frac{(-1)^{\ell}}{\ell!} \partial_L \left(\frac{1}{|x|} \int_{\mathcal{M}} \tilde{\mu}_s(ct-|x|,x') x'^L d^3x'\right)\right.\\
&\left.-\frac{1}{2} \mphi^2 \int d^3 x' \int_{-\infty}^{0} dt' \Theta(t-t'-|x|/c) \tilde{\mu}_s(t',x')\right.\\
&\left.-\frac{1}{2}\mphi^2 \sum_{k=1}^{\infty}\frac{(-1)^{k}}{k!} \partial_{K-1} \left( \frac{x^i}{|x|} \int_{\mathcal{M}}d^3x'  \tilde{\mu}_s(t-|x|/c,x') x'^K \right) \right].
\eal
\ee

\noindent The dependence on $\mathbf{x}$ in the source terms $\tilde{\mu}_s$ are contained in the retarded time. Therefore we can rewrite the derivative to the source as $\partial_L \mu = (-1)^{\ell} c^{-\ell} \mu^{(\ell)}n_L + \mathcal{O}(|x|^{-1})$ where we used $\partial_j |x| = n_j$ with $n_j = \frac{x_j}{|x|}$ and the notation $\mu^{(\ell)}$ denotes the $\ell$ derivative to retarded time. Additionally, we will see the relevant source term in the far zone is already order $\mphi^2$, therefore we give below only the $\mphi^0$ order terms. Expanding to second order in $\ell$ and $k$ and defining the multipole moments
\be
\bal
M_s^{L} &= \int_{\mathcal{M}} \tilde{\mu}_s x'^L d^3x',\\
\eal
\ee
we can write

\be\label{eq:phiNZFZ}
\bal
\varphi=&-\frac{G}{c^4}\bigg( \frac{1}{|x|} M_s +  \frac{n_i}{|x|^2}M_s^{i}-\frac{n_i}{|x| c}\dot{M}_s^{i} -  \frac{\delta^{ij}+3 n^i n^j}{|x|^3}M_s^{ij} - 2  \frac{n^i n^j}{|x|^2 c}\dot{M}_s^{ij} +\\
&\frac{n^i n^j}{|x| c^2}\ddot{M}_s^{ij}  + \mathcal{O}(c^{-4})+ \mathcal{O}(\tilde{m}_{\varphi}^4) \bigg) .
\eal
\ee
In the same way we obtain for $h^{00(0)}$ defining the moments 
\be
M^{00L} = \int_{\mathcal{M}} \mu^{00} x'^L d^3x',
\ee

\be\label{eq:h00NZFZ}
\bal
h^{00} =& \frac{4 G}{c^4}\bigg( \frac{1}{|x|} M^{00} +  \frac{n_i}{|x|^2}M^{00i}-\frac{n_i}{|x| c}\dot{M}^{00i} -  \frac{\delta^{ij}+3 n^i n^j}{|x|^3}M^{00ij} - 2  \frac{n^i n^j}{|x|^2 c}\dot{M}^{00ij} +\\
&\frac{n^i n^j}{|x| c^2}\ddot{M}^{00ij} + \mathcal{O}(c^{-4})+ \mathcal{O}(\tilde{m}_{\varphi}^4)\bigg).
\eal
\ee

\noindent At this stage we will leave the NZ solutions with far zone field points in terms of the moments. When evaluating the far zone contributions we will substitute the source terms \eqref{eq:PNsources}.

\bibliographystyle{ieeetr}
\bibliography{bib}

\end{document}